\documentclass[12pt]{article}

\usepackage{amsfonts}
\usepackage{amssymb}

\newcommand{\smin}{\,\raisebox{0.06em}{${\scriptstyle \in}$}\,}
\newcommand{\nsmin}{\,\raisebox{0.06em}{${\scriptstyle \notin}$}\,}
\newcommand{\ssmin}{\,\raisebox{0.06em}{${\scriptscriptstyle \in}$}\,}
\newcommand{\smsubset}{\,\raisebox{0.06em}{${\scriptstyle \subset}$}\,}

\newtheorem{theorem}{Theorem}[section]

\begin{document}
\title{New Approach to the Integrability \\
       of the Calogero Models}
\author{Michael Forger and Axel Winterhalder
        \thanks{forger@ime.usp.br and winter@ime.usp.br}}
\date{Departamento de Matem\'atica Aplicada, \\
      Instituto de Matem\'atica e Estat\'{\i}stica, \\
      Universidade de S\~ao Paulo, \\
      Caixa Postal 66281, \\[2mm]
      BR--05315-970~ S\~ao Paulo, S.P., Brazil}
\maketitle
\renewcommand{\thefootnote}{\fnsymbol{footnote}}
\footnotetext[1]{Work supported by CNPq (Conselho Nacional de Desenvolvimento
                 Cient\'{\i}fico e Tecno\-l\'o\-gico), Brazil, by FAPESP
                 (Funda\c{c}\~ao de Amparo \`a Pesquisa do Estado de S\~ao
                 Paulo), Brazil, and by DFG (Deutsche Forschungsgemeinschaft),
                 Germany}
\renewcommand{\thefootnote}{\arabic{footnote}}
\setcounter{footnote}{0}
\thispagestyle{empty}
\begin{abstract}        
\noindent
We develop a new, systematic approach towards studying the integrability
of the ordinary Calogero-Moser-Sutherland models as well as the elliptic
Calogero models associated with arbitrary (semi-)simple Lie algebras and
with symmetric pairs of Lie algebras. It is based on the introduction of
a function $F$, defined on the relevant root system and with values in
the respective Cartan subalgebra, satisfying a certain set of combinatoric
identities that ensure, in one stroke, the existence of a Lax representation
and of a dynamical $R$-matrix, given by completely explicit formulas. It is
shown that among the simple Lie algebras, only those belonging to the $A$-%
series admit such a function $F$, whereas the $AIII$-series of symmetric
pairs of Lie algebras, corresponding to the complex Grassmannians
$\, SU(p,q)/S(U(p) \times U(q))$, allows non-trivial solutions when
$\, |p-q| \leq 1$. Apart from reproducing all presently known dynamical
$R$-matrices for Calogero models, our approach provides new ones, namely
for the ordinary models when $\, |p-q| = 1$ \linebreak and for the elliptic
models when $\, |p-q| = 1 \,$ or $\, p = q \,$.
\end{abstract}
\vspace{-5mm}
\begin{flushright}
 \parbox{12em}
 {\begin{center}
   Universidade de S\~ao Paulo \\
   RT-MAP 99/07 \\
   December 1999
  \end{center}}
\end{flushright}

\newpage

\section{Introduction}

The Calogero-Moser-Sutherland models, or Calogero models, for short,
constitute an important class of completely integrable Hamiltonian
systems, intimately related to the theory of (semi-)simple Lie
algebras [1-6]. Unfortunately, the group-theoretical underpinnings
of their integrability are far less understood than in the case of
the Toda models, which allow for a completely general and uniform
treatment in terms of root systems of ordinary (semi-)simple Lie
algebras and/or affine Lie algebras. Indeed, it has proven surprisingly
difficult to extend the results obtained for the ordinary Calogero model
based on $\mathfrak{sl}(n,\mathbb{C})$, such as the Lax pair found in
ref.\ \cite{Mo} and the $R$-matrix given in ref.\ \cite{AT}, to other
simple Lie algebras.

In recent years, a number of attempts has been made to improve this
situation. One of these is based on Hamiltonian reduction, starting
out from the geodesic flow on the corresponding group and leading to
the construction of Lax pairs and dynamical $R$-matrices \cite{ABT}.
However, the approach is rather indirect and the results do not really
crystallize into explicit formulas, except for a few examples that are
worked out ex\-pli\-citly; moreover, it remains unclear why the traditional
more direct approach \cite{OP1,OP2,Pe} works for certain simple Lie
algebras such as $\mathfrak{sl}(n,\mathbb{C})$ but fails for others.
Other authors have addressed important issues concerning integrability
of the elliptic models, where the pair interaction potential is given
by the doubly periodic Weierstra\ss\ function $\wp$ and the main objects
can be extended to include a spectral parameter -- ranging from the
construction of a Lax pair \cite{Kr} and of a dynamical $R$-matrix
\cite{Sk} in the $\mathfrak{sl}(n,\mathbb{C})$-case to the more
recently achieved construction of a Lax pair for other simple Lie
algebras \cite{DP}; however, it seems that dynamical $R$-matrices
for other simple Lie algebras are still missing.

One of the central features that distinguishes the Calogero models
from the (far simpler) Toda models is that their $R$-matrix is not
a numerical object but has a dynamical character. Generally speaking,
the main function of the $R$-matrix is to control the Poisson brackets
between the components of the Lax matrix $L$, according to the well-%
known formula
\begin{equation} \label{eq:PB}
 \{ L_1 , L_2 \}~=~[R_{12},L_1] - [R_{21},L_2]~.
\end{equation}
Numerical $R$-matrices are well understood mathematical objects that
play an important role in the theory of quantum groups. or rather
its classical counterpart, the theory of Lie bialgebras \cite{ST,CP}.
On the other hand, the mathematical status of dynamical $R$-matrices is
not nearly as clear: there does not even seem to exist any generally
accepted definition. The first examples of dynamical $R$-matrices
appeared in the study of the nonlinear sigma models on spheres
\cite{M1,M2,M3} and, more generally, on Riemannian symmetric spaces
\cite{BFLS}. Unfortunately, the analysis of the underlying algebraic
structure is hampered by technical problems due to the fact that
these models are not ultralocal \cite{FT} and hence Poisson brackets
between components of the transition matrix $T$ constructed from the
Lax matrix $L$ show discontinuities that must be removed by regularization.
Therefore, it seems much more promising to undertake such an analysis not in
the context of two-dimensional field theory but in the context of mechanics,
where such technical problems are avoided due to finite-dimensionality of
the phase space. This leads us naturally to the Calogero models -- the
only known class of models with a finite number of degrees of freedom
where dynamical $R$-matrices make their appearance.

In view of this situation, we have developed a new systematic approach towards
studying the integrability of the Calogero models associated with arbitrary
(semi-) \linebreak simple Lie algebras and with symmetric pairs of Lie algebras.
We show that by introducing an appropriate function $F$ defined on the relevant
root system and taking values in the respective Cartan subalgebra, we can
formulate a general Ansatz both for the Lax pair and for the $R$-matrix
through which the proof of integrability (equivalence between the equation
of motion and the Lax equation $\, \dot{L} = [L,A]$) and the proof of the
Poisson bracket relation (\ref{eq:PB}) can be reduced to essentially one
and the same set of combinatorical identities. These state that for any
two roots $\alpha$ and $\beta$
\begin{equation} \label{eq:COMBID1}
 2 \, \mbox{\sl g}_\beta \, \beta(F_\alpha)~
 =~{\mit{\Gamma}}_{\beta,\alpha} \, + \,
   {\mit{\Gamma}}_{\beta,-\alpha}
\end{equation}
(with $\, {\mit{\Gamma}}_{\beta,\pm\alpha} = \mbox{\sl g}_{\beta\pm\alpha}
N_{\beta,\pm\alpha}$) in the case of simple Lie algebras and
\begin{equation} \label{eq:COMBID2}
 2 \, \mbox{\sl g}_\beta \, \beta(F_\alpha)~
 =~{\mit{\Gamma}}_{\beta,\alpha}^\theta \, + \,
   {\mit{\Gamma}}_{\beta,-\alpha}^\theta
\end{equation}
in the case of symmetric pairs; for more details, including an explanation
of the notation, we refer to the main body of the text. It is worth noting
that this is a highly overdetermined system of equations which will admit
non-trivial solutions only in special circumstances; its deeper algebraic
meaning is yet to be discovered.

The plan of the paper is as follows. In Sect.~2, we briefly review the basic
notation and collect the main conventions from the theory of semisimple Lie
algebras and symmetric spaces that are needed in our calculations. Then we
present the construction of the Lax pair and of the $R$-matrix for the
Calogero model, in terms of the function~$F$, for (semi-) simple Lie
algebras in Sect.~2 and for symmetric pairs in Sect.~3, and we show
how both the integrability of the model and the Poisson bracket relation
(\ref{eq:PB}) can be reduced to the combinatorical identity given above.
Moreover, we show in Sect.~2 that the only simple Lie algebras for which
a function $F$ with the desired property exists are those belonging to
the $A$-series, i.e., the Lie algebras $\mathfrak{sl}(n,\mathbb{C})$;
the resulting $R$-matrix coincides with that given in ref.~\cite{AT}
for the ordinary Calogero model and with that given in ref.~\cite{Sk}
for the elliptic Calogero model. In Sect.~3, we study as an example the
$AIII$-series of symmetric pairs of Lie algebras, corresponding to the
complex Grassmannians $\, SU(p,q)/S(U(p) \times U(q))$, and we prove by
explicit calculations that there are non-trivial solutions, depending on
free parameters, when $\, |p-q| \leq 1$. For $\, p = q$, the $R$-matrix thus
found coincides with that given in ref.\ \cite{ABT} for the ordinary Calogero
model, while the one derived for the elliptic Calogero model is new. For
$\, |p-q| = 1$, all $R$-matrices found seem to be new. Together, these
also provide $R$-matrices for all Calogero models, ordinary
and elliptic, associated with any one of the four classical root systems
$A_n$, $B_n$ $C_n$ and $D_n$.



\section{Algebraic Preliminaries}

Root systems belong to the most important and intensively studied objects of
discrete geometry, partly because of the central role they play in various
areas of algebra. The most traditional application of root systems is to
the classification of (semi-)simple Lie algebras and of symmetric pairs
of Lie algebras, due to Cartan. In the first case, one obtains only the
so-called reduced root systems, whereas the second case leads to general
root systems, reduced as well as non-reduced ones, and to the appearance
of non-trivial multiplicities.

Throughout this paper, $\mathfrak{g}$ will always denote a complex semisimple
Lie algebra. Fixing a Cartan subalgebra $\mathfrak{h}$ of $\mathfrak{g}$, let
$\mathfrak{h}^\ast$ denote its dual, $\Delta \smsubset \mathfrak{h}^\ast \,$
the root system of $\mathfrak{g}$ with respect to $\mathfrak{g}$ and
$\mathfrak{h}_{\mathbb{R}}^\ast$ the real subspace of $\mathfrak{h}^\ast$
spanned by the roots $\alpha$ ($\alpha \smin \Delta$). Furthermore, let
$(.\,,.)$ be the invariant bilinear form on $\mathfrak{g}$, normalized
so that with respect to the induced bilinear forms on $\mathfrak{h}$
and $\mathfrak{h}^\ast$, also denoted by $(.\,,.)$ and related to each
other by
\begin{equation} \label{eq:INVBF1}
 (\alpha,\beta)~=~(H_\alpha,H_\beta)~,
\vspace{1mm}
\end{equation}       
with
\begin{equation} \label{eq:HALPH1}
 (H_\alpha,H)~=~\alpha(H) \qquad
 \mbox{for all $\, H \smin \mathfrak{h}$}~,
\vspace{1mm}
\end{equation}
the long roots have length $\sqrt 2$. The real subspace of $\mathfrak{h}$
spanned by the generators $H_\alpha$ ($\alpha \smin \Delta$), on which
$(.\,,.)$ is positive definite, will be denoted by $\mathfrak{h}_%
{\mathbb{R}}$. We also fix a Weyl chamber $C$ and a Weyl alcove $A$;
these are open subsets of $\mathfrak{h}_{\mathbb{R}}$ with boundaries
formed by hyperplanes on which some root $\, \alpha \smin \Delta \,$
vanishes and by hyperplanes on which some root $\, \alpha \smin \Delta \,$
assumes integer values, respectively. Recall that the choice of a Weyl
chamber $C$ establishes an ordering in $\Delta$, the positive roots being
the ones that assume strictly positive values on $C$. The different Weyl
chambers are permuted by the elements of the Weyl group $W(\mathfrak{g})$
of $\mathfrak{g}$, which is the finite group generated by the reflections
$s_\alpha$ in $\mathfrak{h}_{\mathbb{R}}$ along the roots $\, \alpha \smin
\Delta$. Finally we choose a basis of generators $\, \{ E_\alpha \, / \,
\alpha \smin \Delta \} \,$ in $\mathfrak{g}$ normalized so that
\begin{equation} \label{eq:NORM1}
 (E_\alpha,E_\beta)~=~\delta_{\alpha+\beta,0}
\end{equation}
and therefore obeying the standard Cartan-Weyl commutation relations
\begin{eqnarray} \label{eq:CWCR1}
 &[H,E_\alpha]~=~\alpha(H) \, E_\alpha \qquad
 \mbox{for all $\, H \smin \mathfrak{h}$}~,& \label{eq:COMMR1} \\[1mm]
 &[E_\alpha,E_{-\alpha}]~=~H_\alpha~,& \label{eq:CWCR2} \\[1mm]
 &[E_\alpha,E_\beta]~=~N_{\alpha,\beta} \, E_{\alpha+\beta}~,& \label{eq:CWCR3}
\vspace{1mm}
\end{eqnarray}
with structure constants $N_{\alpha,\beta}$ which satisfy $\, N_{\beta,\alpha}
= - N_{\alpha,\beta} \,$ and which, by definition, are supposed to vanish
whenever $\, \alpha + \beta \,$ is not a root -- in particular when
$\, \alpha \pm \beta = 0 \,$. These conditions determine uniquely the
$H_\alpha$ (according to eqn (\ref{eq:HALPH1})) but not the $E_\alpha$,
because it is always possible to rescale the root generators $E_\alpha$
according to
\begin{equation}
 E_\alpha~~\rightarrow~~E_\alpha^\prime~=~a_\alpha \, E_\alpha \qquad
 \mbox{with $\, a_\alpha a_{-\alpha} = 1$}
\end{equation}
which entails a rescaling of the structure constants according to
\begin{equation}
 N_{\alpha,\beta}~~\rightarrow~~
 N_{\alpha,\beta}^\prime~=~{a_\alpha a_\beta \over a_{\alpha+\beta}} \,
                           N_{\alpha,\beta}~.
\end{equation}
This freedom can be used in order to impose the additional normalization
conditions
\begin{equation} \label{eq:NORM2}
 N_{-\alpha,-\beta}~=~- \, N_{\alpha,\beta}~.
\end{equation}
As it it turns out, this determines the root generators $E_\alpha$ and the
structure constants $N_{\alpha,\beta}$ uniquely up to signs. Another important
relation between the structure constants that we shall use frequently is the
following cyclic identity: if $\, \alpha,\beta,\gamma \smin \Delta \,$ are
any three roots that add up to zero, then
\begin{equation} \label{eq:CYCID}
 N_{\alpha,\beta}~=~N_{\beta,\gamma}~=~N_{\gamma,\alpha}~.
\end{equation}
In addition, assuming that $r$ is the rank of $\mathfrak{g}$, we may
choose an orthonormal basis $\, \{ H_1 , \ldots , H_r \} \,$ of
$\mathfrak{h}_{\mathbb{R}}$; then
\begin{equation} \label{eq:RTIDE1}
 \sum_{j=1}^r \alpha(H_j) H_j~=~H_{\alpha}~.
\end{equation}
In the subsequent computations we shall make extensive use of these relations,
often without further mention. For proofs, see for example \cite{He}.

In the symmetric space situation, we shall assume more specifically that
$\mathfrak{g}$ is the complexification of a real semisimple Lie algebra
$\mathfrak{g}_0$ of rank $r$ and that we are given a direct decomposition
\begin{equation} \label{eq:CARDEC1}
 \mathfrak{g}~=~\mathfrak{k} \oplus \mathfrak{m}
\end{equation}
of $\mathfrak{g}$ induced from a Cartan decomposition
\begin{equation} \label{eq:CARDEC2}
 \mathfrak{g}_0~=~\mathfrak{k}_0 \oplus \mathfrak{m}_0
\end{equation}
of $\mathfrak{g}_0$; then
\begin{equation} \label{eq:CARDEC3}
 \mathfrak{g}_k~=~\mathfrak{k}_0 \oplus i \mathfrak{m}_0
\end{equation}
defines a compact real form $\mathfrak{g}_k$ of $\mathfrak{g}$.
Writing $\sigma$ for the conjugation in $\mathfrak{g}$ with respect to
$\mathfrak{g}_0$ and $\tau$ for the conjugation in $\mathfrak{g}$ with
respect to $\mathfrak{g}_k$ (note that these commute), the corresponding
Cartan involution $\theta$ is simply their product: $\theta = \sigma \tau
= \tau \sigma$. This is an involutive automorphism of $\mathfrak{g}$ which
is $+1$ on $\mathfrak{k}$ and $-1$ on $\mathfrak{m}$; moreover, it preserves
the invariant bilinear form $(.\,,.)$ on $\mathfrak{g}$ introduced before,
which means that the direct sums in eqns (\ref{eq:CARDEC1})-(\ref{eq:CARDEC3})
are orthogonal, characterizing the direct decompositions (\ref{eq:CARDEC2})
and (\ref{eq:CARDEC3}) as being associated with a Riemannian symmetric
space of the noncompact type and of the compact type, respectively.
The decomposition of elements corresponding to the direct decompositions
(\ref{eq:CARDEC1})-(\ref{eq:CARDEC3}) into eigenspaces under $\theta$ is
written as
\begin{equation} \label{eq:CARDEC4}
 X~=~X_{\mathfrak{k}} + X_{\mathfrak{m}}~,
\end{equation}
where obviously
\begin{equation} \label{eq:CARDEC5}
 X_{\mathfrak{k}}~=~{\textstyle {1 \over 2}} \, (X + \theta X)~~,~~
 X_{\mathfrak{m}}~=~{\textstyle {1 \over 2}} \, (X - \theta X)~.
\end{equation}
Next, let $\mathfrak{a}_0$ denote a maximal abelian subalgebra of
$\mathfrak{m}_0$, $\mathfrak{h}_0$ a maximal abelian subalgebra of
$\mathfrak{g}_0$ with $\, \mathfrak{a}_0 \smsubset \mathfrak{h}_0$,
$\mathfrak{b}_0$ its orthogonal complement in $\mathfrak{h}_0 \,$
and let $\mathfrak{a}$, $\mathfrak{b}$ and $\mathfrak{h}$ denote
the corresponding complexifications; then $\mathfrak{h}$ is a
Cartan subalgebra of $\mathfrak{g}$ and
\begin{equation} \label{eq:CARDEC6}
\begin{array}{ccccccc}
 \mathfrak{h}~=~\mathfrak{b} \, \oplus \, \mathfrak{a} &
 \quad & \mbox{with} & &
 \mathfrak{b}~=~\mathfrak{h} \cap \mathfrak{k} &,&
 \mathfrak{a}~=~\mathfrak{h} \cap \mathfrak{m}~, \\[1mm]
 \mathfrak{h}_0~=~\mathfrak{b}_0 \, \oplus \, \mathfrak{a}_0 &
 \quad & \mbox{with} & &
 \mathfrak{b}_0~=~\mathfrak{h} \cap \mathfrak{k}_0 &,&
 \mathfrak{a}_0~=~\mathfrak{h} \cap \mathfrak{m}_0~.
\end{array}
\end{equation}  
The real subspace $\mathfrak{h}_{\mathbb{R}}$ of $\mathfrak{h}$ introduced
before then splits according to
\begin{equation} \label{eq:CARDEC7}
 \mathfrak{h}_{\mathbb{R}}~
 =~i \mathfrak{b}_0 \, \oplus \, \mathfrak{a}_0
\end{equation}
and the decomposition of elements corresponding to the direct decompositions
(\ref{eq:CARDEC6})-(\ref{eq:CARDEC7}) into eigenspaces under $\theta$ is
written as
\begin{equation} \label{eq:CARDEC8}
 H~=~H_{\mathfrak{b}} + H_{\mathfrak{a}}~,
\end{equation}
where obviously
\begin{equation} \label{eq:CARDEC9}
 H_{\mathfrak{b}}~=~{\textstyle {1 \over 2}} \, (H + \theta H)~~,~~
 H_{\mathfrak{a}}~=~{\textstyle {1 \over 2}} \, (H - \theta H)~.
\end{equation}
Next, we note that the conjugations $\sigma$, $\tau$ and the involution
$\theta$ induce bijective transformations of the root system $\Delta$ onto
itself that for the sake of simplicity will again be denoted by $\sigma$,
$\tau$ and $\theta$, respectively; they are characterized by the condition
$$
 \sigma(\mathfrak{g}^\alpha)~=~\mathfrak{g}^{\sigma\alpha}~~,~~
 \tau  (\mathfrak{g}^\alpha)~=~\mathfrak{g}^{\tau  \alpha}~~,~~
 \theta(\mathfrak{g}^\alpha)~=~\mathfrak{g}^{\theta\alpha}~,
$$
where $\mathfrak{g}^\alpha$ is the one-dimensional subspace of $\mathfrak{g}$
generated by $E_\alpha$ and are explicitly given by
\begin{equation} \label{eq:SKONJ1}
 (\sigma\alpha)(H)~=~\overline{\alpha(\sigma H)}~~,~~
 (\tau  \alpha)(H)~=~\overline{\alpha(\tau   H)}~~,~~
 (\theta\alpha)(H)~=~          \alpha(\theta H) \qquad
 \mbox{for $\, H \smin \mathfrak{h}$}~.
\end{equation}
In the following we write $\bar{\alpha}$ for the linear functional
on $\mathfrak{a}$ obtained by restricting a given linear functional
$\alpha$ on $\mathfrak{h}$ to $\mathfrak{a}$. In this way the root
system $\, \Delta \smsubset \mathfrak{h}^\ast \,$ gives rise to the
restricted root system $\, \bar{\Delta} \smsubset \mathfrak{a}^\ast$:
\begin{equation}
 \bar{\Delta}~=~\{ \, \bar{\alpha} = \alpha\vert_{\mathfrak{a}}^{} \, / \,
                   \alpha \smin \Delta \, \}~.
\end{equation}
Accordingly the root system $\Delta$ itself decomposes into two parts,
\begin{equation} \label{eq:SDECRS}
 \Delta~=~\tilde{\Delta} \cup \Delta_0~,
\vspace{-1mm}
\end{equation}
where
\vspace{-2mm}
\begin{eqnarray}
 \tilde{\Delta} \!\!
 &=~     \{ \, \alpha \smin \Delta \, / \, \theta\alpha \neq \alpha \, \}~
  =&\!\! \{ \, \alpha \smin \Delta \, / \, \bar{\alpha} \neq 0 \, \}~, \\[1mm]
 \Delta_0 \!\!
 &=~     \{ \, \alpha \smin \Delta \, / \, \theta\alpha = \alpha \, \}~
  =&\!\! \{ \, \alpha \smin \Delta \, / \, \bar{\alpha} = 0 \, \}~,
\end{eqnarray}
where obviously $\Delta_0$ is the root system of the subalgebra $\mathfrak{k}$
since linear functionals on $\mathfrak{h}$ whose restriction to $\mathfrak{a}$
vanishes may be naturally identified with linear functionals on $\mathfrak{b}$.
In contrast to ordinary roots, restricted roots will in general have non-%
trivial multiplicities, defined as follows:
\begin{equation} \label{eq:MULTI1}
 m_{\lambda}~=~\mbox{card} \, \{ \, \alpha \smin \Delta \, / \,
                                    \bar{\alpha} = \lambda \, \}~.
\end{equation}
As before, we also fix a Weyl chamber $C$ and a Weyl alcove $A$, but these
are now open subsets of $\mathfrak{a}_0$ (rather than of $\mathfrak{h}_%
{\mathbb{R}}$) with boundaries formed by hyperplanes on which some root
$\, \alpha \smin \Delta \,$ vanishes and by hyperplanes on which some
root $\, \alpha \smin \Delta \,$ assumes integer values, respectively.
The choice of such a Weyl chamber $C$ now establishes an ordering only
in $\tilde{\Delta}$, the positive roots in $\tilde{\Delta}$ being the
ones that assume strictly positive values on $C$, but this can of course
be extended to an ordering in $\Delta$. It is interesting and useful
to note the behavior of the conjugations $\sigma$, $\tau$ and of the
involution $\theta$ with respect to this ordering:
$$
\begin{array}{c}
 \alpha \smin \tilde{\Delta}^\pm~~\Longrightarrow~~
 \sigma\alpha \smin \tilde{\Delta}^\pm~,~\tau\alpha~=~- \, \alpha~,~
 \theta\alpha \smin \tilde{\Delta}^\mp~, \\[2mm]
 \alpha \smin \Delta_0^{}~~\Longrightarrow~~
 \sigma\alpha~=~- \, \alpha~,~\tau\alpha~=~- \, \alpha~,~
 \theta\alpha~=~\alpha~.
\end{array}
$$
Again, the different Weyl chambers are permuted by the elements of the
Weyl group $W(\mathfrak{g},\theta)$ of $(\mathfrak{g},\theta)$, which is
the quotient of the subgroup $W_\theta(\mathfrak{g})$ of the Weyl group
$W(\mathfrak{g})$ of $\mathfrak{g}$ consisting of those elements that
commute with the involution $\theta$, modulo those elements that act
trivially on $\mathfrak{a}_0$. Finally we assume the basis of generators
$\, \{ E_\alpha \, / \, \alpha \smin \Delta \} \,$ in $\mathfrak{g}$ to be
chosen so that, apart from the normalization condition (\ref{eq:NORM1}),
the Cartan-Weyl commutation relations (\ref{eq:CWCR1}), the additional
normalization condition (\ref{eq:NORM2}) and the cyclic identity
(\ref{eq:CYCID}), we have the following simple behavior under the
conjugations $\sigma$, $\tau$ and the involution $\theta$:
\begin{eqnarray} \label{eq:NORM3}
 &\sigma E_\alpha~=~- \, E_{\sigma\alpha}~,& \nonumber \\[1mm]
 &\tau E_\alpha~=~- \, E_{-\alpha}~,& \\[1mm]
 &\theta E_\alpha~=~E_{\theta\alpha}~.& \nonumber
\end{eqnarray}
In particular, the structure constants are $\theta$-invariant:
\begin{equation} \label{eq:SINVSC}
 N_{\theta\alpha,\theta\beta}~=~N_{\alpha,\beta}~.
\end{equation}
That such a choice is always possible is a non-trivial statement which
apparently cannot be easily found in the literature and whose proof we
therefore present in Appendix 1. \linebreak In addition, assuming that
$r$ is the rank of the symmetric pair $(\mathfrak{g},\theta)$ and $r+s$
is the rank of the Lie algebra $\mathfrak{g}$ (which by definition means
that $\, r = {\rm dim} \, \mathfrak{a} \,$ and $\, r+s = {\rm dim} \,
\mathfrak{h}$), we may also choose an orthonormal basis $\, \{ H_1 ,
\ldots , H_r \} \,$ of $\mathfrak{a}_0$ together with an orthonormal
basis $\, \{ H_{r+1} , \ldots , H_{r+s} \} \,$ of $i\mathfrak{b}_0$
to form an orthonormal basis $\, \{ H_1 , \ldots , H_{r+s} \} \,$ of
$\mathfrak{h}_{\mathbb{R}}$; then
\begin{equation} \label{eq:RTIDE2}
 \sum_{j=1}^r \alpha(H_j) H_j~=~(H_{\alpha})_{\mathfrak{a}}~~,~~
 \sum_{j=r+1}^{r+s} \alpha(H_j) H_j~=~(H_{\alpha})_{\mathfrak{b}}~.
\end{equation}

\section{Calogero Models for Semisimple Lie Algebras}

\subsection{Definition of the Models, Lax Pairs and R-Matrices}

With the conventions stated above we can introduce the Calogero model
associated with the root system $\Delta$ of the Lie algebra $\mathfrak{g}$.
To begin with, we have to specify the configuration space of the theory,
which in the case of the ordinary (rational or trigonometric) Calogero
model is the previously mentioned Weyl chamber $C$ and in the case of
the elliptic Calogero model is the previously mentioned Weyl alcove $A$.
In any case, $Q$ is an open subset of $\mathfrak{h}_{\mathbb{R}}$, so
the phase space of the model is the associated cotangent bundle $T^* Q
= Q \times \mathfrak{h}_{\mathbb{R}}^*$, which for the sake of simplicity
will often be identified with the tangent bundle $\, TQ = Q \times
\mathfrak{h}_{\mathbb{R}}$. The Hamiltonian for the ordinary (rational
or trigonometric) Calogero model is defined by
\begin{equation} \label{eq:OHAMF1}
 H(q,p)~=~{1 \over 2} \, \Bigl( \, \sum_{j=1}^r \, p_j^2 \, + \,
          \sum_{\alpha \ssmin \Delta} \,
          \mbox{\sl g}_\alpha^2 \, w(\alpha(q))^2 \, \Bigr)~,
\end{equation}
while that for the elliptic Calogero model is defined by
\begin{equation} \label{eq:EHAMF1}
 H(q,p)~=~{1 \over 2} \, \Bigl( \, \sum_{j=1}^r \, p_j^2 \, + \,
          \sum_{\alpha \ssmin \Delta} \,
          \mbox{\sl g}_\alpha^2 \, \wp \>\! (\alpha(q)) \, \Bigr)~,
\end{equation} 
where $w$ is a smooth, real-valued function on $\mathbb{R} \backslash \{0\}$
to be specified soon, while $\wp$ denotes the doubly periodic Weierstra\ss\
function (see below). The coefficients $\mbox{\sl g}_\alpha$ ($\alpha \smin
\Delta$) are positive real coupling constants satisfying the condition
\begin{equation} \label{eq:COUPC1}
 \mbox{\sl g}_{-\alpha}~=~\mbox{\sl g}_\alpha~.
\end{equation}
A stronger assumption that we shall always make is that they are invariant
under the action of the Weyl group $W(\mathfrak{g})$ of $\mathfrak{g}$:
\begin{equation} \label{eq:COUPC2}
 \mbox{\sl g}_{w\alpha}~=~\mbox{\sl g}_\alpha \qquad
 \mbox{for all $\, w \smin W(\mathfrak{g})$}~.
\end{equation}
For later use we introduce the following combination of coupling constants
and structure constants:
\begin{equation} \label{eq:GAMMA}
 {\mit{\Gamma}}_{\alpha,\beta}~
 =~\mbox{\sl g}_{\alpha + \beta} \, N_{\alpha,\beta}~.
\end{equation}
Using the abbreviations
\begin{equation} \label{eq:ORTIMP}
 p~=~\sum_{j=1}^r \, p_j H_j~~,~~q~=~\sum_{j=1}^r \, q_j H_j~,
\end{equation} 
we can write the Hamiltonian equations of motion as vector equations in
$\mathfrak{h}_{\mathbb{R}}$: they read
\begin{equation} \label{eq:OHAMEQ}
 \dot{q}~=~p~~~,~~~
 \dot{p}~=~- \, \sum_{\alpha \ssmin \Delta} \mbox{\sl g}_\alpha^2 \;
                w(\alpha(q)) \, w^\prime(\alpha(q)) \; H_\alpha
\end{equation}
for the ordinary (rational or trigonometric) Calogero model and
\begin{equation} \label{eq:EHAMEQ}
 \dot{q}~=~p~~~,~~~
 \dot{p}~=~- \; {\textstyle {1 \over 2}} \,
               \sum_{\alpha \ssmin \Delta} \mbox{\sl g}_\alpha^2 \;
                \wp^\prime(\alpha(q)) \; H_\alpha
\end{equation}  
for the elliptic Calogero model.

Concerning the choice of the potential functions $w$ in eqn (\ref{eq:OHAMF1})
and $\wp$ in eqn (\ref{eq:EHAMF1}), it is well known that these functions must
satisfy a certain set of functional equations that we shall discuss briefly in
order to make our presentation self-contained. An elementary property is that
$w$ is supposed to be odd while $\wp$ is even:
\begin{equation} \label{eq:POTF1}
 w(-t)~=~- \, w(t)~~,~~\wp \>\! (-t)~=~\wp \>\! (t)~.
\end{equation}
The basic functional equation imposed on $w$ is the following:
\begin{equation} \label{eq:OFUNEQ1}
 \Bigl( \, {w^\prime(s) \over w(s)} \, + \,
           {w^\prime(t) \over w(t)} \, \Bigr) \, w(s+t) \,
 + \, w(s) \, w(t)~=~0~.
\end{equation}
Differentiating with respect to $s$ and to $t$ and subtracting the results
gives a second functional equation:
\begin{equation} \label{eq:OFUNEQ2}
 \Bigl( \, {w^{\prime\prime}(s) \over 2 \, w(s)} \, - \,
           {w^{\prime\prime}(t) \over 2 \, w(t)} \, \Bigr) \, w(s+t)~
 =~w(s) \, w^\prime(t) \, - \, w^\prime(s) \, w(t)~.
\end{equation}
The solutions are derived in refs \cite{OP1,OP2,Pe}: there are essentially
three different ones, satisfying the relation
\begin{equation} \label{eq:OFUNEQ3}
 {w^{\prime\prime}(t) \over 2\,w(t)} \, - \, w(t)^2~=~- \, {k \over 2}~,
\end{equation}  
where $k$ is a numerical constant that allows to distinguish between them,
namely
\begin{equation} \label{eq:OFUNEQ4}
 \begin{array}{cccc}
  w(t)~=~1/\,t & \quad & \mbox{with} & k = 0~, \\[2mm]
  w(t)~=~1/\sin t & \quad & \mbox{with} & k = +1~, \\[2mm]
  w(t)~=~1/\sinh t & \quad & \mbox{with} & k = -1~,
 \end{array}
\end{equation}
which explains the terminology ``rational'' (for the first case) and
``trigonometric'' (for the last two cases). On the other hand, $\wp$
is a meromorphic function on the complex plane with second order poles
at the points of its lattice of periodicity $\Lambda$, consisting of
the integer linear combinations $\, n_1 \omega_1 + n_2 \omega_2 \,$
of its two basic periods $\omega_1$ and $\omega_2$, which are two
arbitrarily chosen but fixed complex numbers; explicitly, it can
be defined by the series expansion
$$
 \wp \>\! (z)~
 =~{1 \over z^2} \sum_{{n_1,n_2 \ssmin \mathbb{Z}} \atop
                       {(n_1,n_2) \neq (0,0)}}
   \left( {1 \over (z - n_1 \omega_1 - n_2 \omega_2)^2} \, - \,
          {1 \over (n_1 \omega_1 + n_2 \omega_2)^2} \right)\!~.
$$
For later use, it will be convenient to also introduce the Weierstra\ss\
sigma function, an entire function on the complex plane that can be defined
by the infinite product expansion
$$
 \sigma(z)~
 =~z \prod_{{n_1,n_2 \ssmin \mathbb{Z}} \atop {(n_1,n_2) \neq (0,0)}}
     \left( 1 \, - \, {z \over n_1 \omega_1 + n_2 \omega_2} \right)
     \exp \left\{ {z \over n_1 \omega_1 + n_2 \omega_2} \, + \, {1 \over 2} \,
                  {z^2 \over (n_1 \omega_1 + n_2 \omega_2)^2} \right\}
$$
and the Weierstra\ss\ zeta function, another meromorphic function on the
complex plane with first order poles at the points of the same lattice of
periodicity, which can be defined by the series expansion
$$
 \zeta(z)~
 =~{1 \over z} \sum_{{n_1,n_2 \ssmin \mathbb{Z}} \atop
                     {(n_1,n_2) \neq (0,0)}}
               \left( {1 \over z - n_1 \omega_1 - n_2 \omega_2} \, + \,
                      {z + n_1 \omega_1 + n_2 \omega_2 \over
                       n_1 \omega_1 + n_2 \omega_2} \right)\!~.
$$
Obviously,
\begin{equation}
 \zeta(z)~=~{\sigma^\prime(z) \over \sigma(z)}~~,~~
 \wp \>\! (z)~=~- \, \zeta^\prime(z)~.
\end{equation}
Finally, we set
\begin{equation}
 \Phi(z_1,z_2)~=~{\sigma(z_1+z_2) \over \sigma(z_1) \>\! \sigma(z_2)}~.
\end{equation}
Clearly, these functions have the following symmetry properties under
reflection at the origin:
\begin{equation}
 \sigma(-z)~=~- \, \sigma(z)~~,~~
 \zeta(-z)~=~- \, \zeta(z)~~,~~
 \wp \>\! (-z)~=~\wp \>\! (z)~,
\end{equation}
\begin{equation} \label{eq:EFUNEQ1}
 \Phi(-z_1,-z_2)~=~- \Phi(z_1,z_2)~.
\end{equation}
Moreover, they satisfy a set of functional equations that can all be derived
from a single quartic identity for the Weierstra\ss\ sigma function, namely
\begin{eqnarray*}
\lefteqn{\sigma(z_1 + z_2) \, \sigma(z_1 - z_2) \,
         \sigma(z_3 + z_4) \, \sigma(z_3 - z_4)}                             \\
 &+&\!\!\! \sigma(z_2 + z_3) \, \sigma(z_2 - z_3) \,
           \sigma(z_1 + z_4) \, \sigma(z_1 - z_4)                            \\
 &+&\!\!\! \sigma(z_3 + z_1) \, \sigma(z_3 - z_1) \,
           \sigma(z_2 + z_4) \, \sigma(z_2 - z_4)~=~0~.
\end{eqnarray*}
We shall restrict ourselves to listing the identities that will actually be
needed in the following calculations: they are
\begin{eqnarray}
 &\Phi(s,u) \, \Phi(-s,u)~=~- \, \left( \wp \>\! (s) - \wp \>\! (u) \right)~,&
 \label{eq:EFUNEQ2} \\[1mm]
 &\Phi(s,u) \, \Phi^\prime(-s,u) \, - \, \Phi^\prime(s,u) \, \Phi(-s,u)~
 =~\wp^\prime(s)~,&
 \label{eq:EFUNEQ3} \\[1mm]
 &\Phi(s,u) \, \Phi^\prime(t,u) \, - \, \Phi^\prime(s,u) \, \Phi(t,u)~
 =~\!\left( \wp \>\! (s) - \wp \>\! (t) \right) \Phi(s+t,u)~,&
 \label{eq:EFUNEQ4} \\[1mm]
 &\Phi(-s,v-u) \, \Phi(s+t,v) \, + \, \Phi(-t,u-v) \, \Phi(s+t,u)~
 =~- \, \Phi(s,u) \, \Phi(t,v)~,&
 \label{eq:EFUNEQ5} \\[1mm]
 &\Phi(-s,u-v) \, \Phi(s,u) \, + \left( \zeta(v-u) + \zeta(u) \right) \Phi(s,v)~
 =~\Phi^\prime(s,v)~,&
 \label{eq:EFUNEQ6}
\end{eqnarray}
where $\Phi^\prime$ denotes the derivative of $\Phi$ with respect to the
first argument and all arguments must be different from zero mod $\Lambda$.
Note that, for example, eqn (\ref{eq:EFUNEQ3}) can be immediately derived
from eqn (\ref{eq:EFUNEQ2}) by differentiation with respect to the first
argument and from eqn ({\ref{eq:EFUNEQ4}) by passing to the limit $\, s+t
\rightarrow 0$. Note also that these properties do not depend on the
choice of the periods $\omega_1$ and $\omega_2$, but for the applications
we have in mind, it will be important to suppose that $\omega_1$ equals $1$
whereas $\omega_2$ is a complex parameter, usually denoted by $\tau$.

With these preliminaries out of the way, we can pass to discuss the Lax
pair and the $R$-matrix of the Calogero model. It is at this point that
an important difference between the ordinary and the elliptic Calogero
models appears, because the Lax pair and the $R$-matrix of the latter
depend on spectral parameters that are absent in the former. However,
these do not enter through the root system, as happens in the Toda
models, but rather show up explicitly in the coefficient functions.

Within the present framework, the Lax pair consists of two mappings%
\footnote{As we shall see, $A$ and $R$ depend only on the position
variables and not on the momentum variables, i.e., they are functions
on $T^* Q$ that descend to functions on $Q$.}
\addtocounter{footnote}{-1}
\begin{equation} \label{eq:LAXPM}
 L : T^* Q \longrightarrow \mathfrak{g}
 \qquad \mbox{and} \qquad
 A : Q \longrightarrow \mathfrak{g}
\end{equation}
each of which will in the elliptic case depend on an additional spectral
parameter $u$, such that the Hamiltonian equations of motion can be rewritten
in the Lax form, namely
\begin{equation} \label{eq:OLAXEQ}
 \dot{L}~=~[L,A]
\end{equation}
for the ordinary (rational or trigonometric) Calogero model and
\begin{equation} \label{eq:ELAXEQ}
 \dot{L}(u)~=~[L(u),A(u)]
\end{equation}
for the elliptic Calogero model. Similarly, the dynamical $R$-matrix is a
mapping\footnotemark
\begin{equation} \label{eq:RMATM}
 R : Q \longrightarrow \mathfrak{g} \otimes \mathfrak{g}
\end{equation}
which will in the elliptic case depend on two additional spectral parameters
$u$ and $v$, such that the Poisson brackets of $L$ can be written in the form
\begin{equation} \label{eq:ORMAT1}
 \{ L_1 , L_2 \}~=~[R_{12},L_1] - [R_{21},L_2]
\end{equation}
for the ordinary (rational or trigonometric) Calogero model and
\begin{equation} \label{eq:ERMAT1}
 \{L_1(u),L_2(v)\}~=~[R_{12}(u,v),L_1(u)] - [R_{21}(v,u),L_2(v)]
\end{equation}
for the elliptic Calogero model, where as usual $\, L_1 = L \otimes 1 \,,\,
L_2 = 1 \otimes L \,$ and $\, R_{12} = R \,,\linebreak R_{21} = R^T$, with
$.^T$ denoting transposition of the two factors in the tensor product.
\linebreak
As is well known, the Lax equations (\ref{eq:OLAXEQ}) or (\ref{eq:ELAXEQ})
imply that the $\mbox{ad}(\mathfrak{g})$-invariant polynomials on
$\mathfrak{g}$ are conserved under the Hamiltonian flow, whereas the
Poisson bracket relations (\ref{eq:ORMAT1}) or (\ref{eq:ERMAT1}) imply
that they are pairwise in involution.

In analogy with the formulas found in the literature for special cases, in
particular the case where $\, \mathfrak{g} = \mathfrak{sl}(r+1,\mathbb{C})$
\cite{AT,Sk}, we postulate a Lax pair and an $R$-matrix; they are explicitly
given by
\begin{equation} \label{eq:OLAXPL}
 L~=~\sum_{j=1}^r \, p_j H_j \, + \,
     \sum_{\alpha \ssmin \Delta} i \, \mbox{\sl g}_\alpha \,
     w(\alpha(q)) \, E_\alpha~,
\end{equation}
\begin{equation} \label{eq:OLAXPA1}
 A~=~- \, \sum_{\alpha \ssmin \Delta} i \, \mbox{\sl g}_\alpha \,
          {w^{\prime\prime}(\alpha(q)) \over 2\,w(\alpha(q))} \, F_\alpha \,
     + \, \sum_{\alpha \ssmin \Delta} i \, \mbox{\sl g}_\alpha \,
          w^\prime(\alpha(q)) \, E_\alpha~,
\vspace{1mm}
\end{equation}
\begin{equation} \label{eq:ORMAT2}
 R~=~\sum_{\alpha \ssmin \Delta} w(\alpha(q)) \,
     F_\alpha \otimes E_\alpha \, + \,
     \sum_{\alpha \ssmin \Delta} {w^\prime(\alpha(q)) \over w(\alpha(q))} \,
     E_\alpha \otimes E_{-\alpha}
\vspace{2mm}
\end{equation}
for the ordinary (rational or trigonometric) Calogero model and by
\begin{equation} \label{eq:ELAXPL}
 L(u)~=~\sum_{j=1}^r \, p_j H_j \, + \,
        \sum_{\alpha \ssmin \Delta} i \, \mbox{\sl g}_\alpha \,
        \Phi(\alpha(q),u) \, E_\alpha~,
\end{equation}
\begin{equation} \label{eq:ELAXPA1}
 A(u)~=~- \, \sum_{\alpha \ssmin \Delta} i \, \mbox{\sl g}_\alpha \,
             \wp \>\! (\alpha(q)) \, F_\alpha \,
        + \, \sum_{\alpha \ssmin \Delta} i \, \mbox{\sl g}_\alpha \,
             \Phi^\prime(\alpha(q),u) \, E_\alpha~,
\vspace{-3mm}
\end{equation}
\begin{eqnarray} \label{eq:ERMAT2}
 R(u,v) \!\!
 &=&\!\! - \, \sum_{j=1}^r \, (\zeta(u-v) + \zeta(v)) \, H_j \otimes H_j
                                                                   \nonumber \\
 & &\!\! + \, \sum_{\alpha \ssmin \Delta} \Phi(\alpha(q),v) \,
              F_\alpha \otimes E_\alpha                                      \\
 & &\!\! - \, \sum_{\alpha \ssmin \Delta} \Phi(\alpha(q),u-v) \,
              E_\alpha \otimes E_{-\alpha}                         \nonumber
\end{eqnarray}
for the elliptic Calogero model. Here, the $F_\alpha$ form a collection of
generators which are supposed to belong to $\mathfrak{h}_{\mathbb{R}}$ and
can be viewed as a (vector valued) function
\begin{equation} \label{eq:FUNF1}
 F: \Delta~\longrightarrow~\mathfrak{h}_{\mathbb{R}}
\end{equation}
which we shall suppose to be even:
\begin{equation} \label{eq:FUNF2}
 F_{-\alpha}~=~F_\alpha~.
\end{equation}
In the sequel we shall derive a combinatoric equation which determines
$F$ completely. Together with the functional equations discussed above,
it will be the basic ingredient for proving the equivalence between the
Hamiltonian equations of motion and the Lax equations, as well as the
validity of the Poisson bracket relations.

For the elliptic Calogero model, the calculation goes as follows. First,
we use the eqns (\ref{eq:ELAXPL}) and (\ref{eq:ELAXPA1}) to compute the
following expression for the difference between the two sides of the Lax
equation (\ref{eq:ELAXEQ}):
\begin{eqnarray*}
 \dot{L}(u) \!\!\!&-&\!\!\! [L(u),A(u)]                                 \\[3mm]
 &=&\!\! \dot{p} \,
         + \, \sum_{\alpha \ssmin \Delta} i \, \mbox{\sl g}_\alpha \,
              \Phi^\prime(\alpha(q),u) \, \alpha(\dot{q}) \, E_\alpha \,
         - \, \sum_{\alpha \ssmin \Delta} i \, \mbox{\sl g}_\alpha \,
              \Phi^\prime(\alpha(q),u) \, \alpha(p) \, E_\alpha              \\
 & &\!\! + \, \sum_{\alpha,\gamma \ssmin \Delta}
              \mbox{\sl g}_\alpha \, \mbox{\sl g}_\gamma \, \wp(\alpha(q)) \,
              \Phi(\gamma(q),u) \; \gamma(F_\alpha) \, E_\gamma              \\
 & &\!\! + \; {\textstyle {1 \over 2}} \,
              \sum_{\alpha \ssmin \Delta} \mbox{\sl g}_\alpha^2 \,
              \Bigl( \Phi(\alpha(q),u) \, \Phi^\prime(-\alpha(q),u) \, - \,
                     \Phi(-\alpha(q),u) \, \Phi^\prime(\alpha(q),u) \Bigr)
              H_\alpha                                                       \\
 & &\!\! + \; {\textstyle {1 \over 2}} \,
              \sum_{\alpha,\beta,\gamma \ssmin \Delta \atop
                    \alpha + \beta = \gamma}
              \mbox{\sl g}_\alpha \, \mbox{\sl g}_\beta \,
              \Bigl( \Phi(\alpha(q),u) \, \Phi^\prime(\beta(q),u) \, - \,
                     \Phi^\prime(\alpha(q),u) \, \Phi(\beta(q),u) \Bigr) \,
              N_{\alpha,\beta} \, E_\gamma~.
\end{eqnarray*}
Here, the double sum over commutators $\, [E_\alpha,E_\beta] \,$ appearing in
the commutator between $L(u)$ and $A(u)$ has, after antisymmetrization of the
coefficients in $\alpha$ and $\beta$, been split into two contributions, one
corresponding to terms where $\, \alpha + \beta = 0 \,$ and the other to terms
where $\, \alpha + \beta \neq 0$, which are nonzero only if $\, \alpha + \beta
\smin \Delta$. Inserting the functional equations (\ref{eq:EFUNEQ3}),
(\ref{eq:EFUNEQ4}) and rearranging terms, we get
\begin{eqnarray*}
 \dot{L}(u) \!\!\!&-&\!\!\! [L(u),A(u)]                                 \\[3mm]
 &=&\!\! \dot{p} \,
         + \; {\textstyle {1 \over 2}} \,
              \sum_{\alpha \ssmin \Delta} \mbox{\sl g}_\alpha^2 \;
              \wp^\prime(\alpha(q)) \; H_\alpha \,
         + \, \sum_{\alpha \ssmin \Delta} i \, \mbox{\sl g}_\alpha \,
              \Phi^\prime(\alpha(q),u) \, \alpha(\dot{q} - p) \, E_\alpha    \\
 & &\!\! + \, \sum_{\alpha,\gamma \ssmin \Delta}
              \mbox{\sl g}_\alpha \, \mbox{\sl g}_\gamma \, \wp(\alpha(q)) \,
              \Phi(\gamma(q),u) \; \gamma(F_\alpha) \, E_\gamma              \\
 & &\!\! + \; {\textstyle {1 \over 2}} \,
              \sum_{\alpha,\gamma \ssmin \Delta \atop
                    \gamma - \alpha \ssmin \Delta}
              \mbox{\sl g}_\alpha \, \mbox{\sl g}_{\gamma-\alpha} \,
              \wp(\alpha(q)) \, \Phi(\gamma(q),u) \;
              N_{\alpha,\gamma-\alpha} \, E_\gamma                           \\
 & &\!\! - \; {\textstyle {1 \over 2}} \,
              \sum_{\beta,\gamma \ssmin \Delta \atop
                    \gamma - \beta \ssmin \Delta}
              \mbox{\sl g}_{\gamma-\beta} \, \mbox{\sl g}_\beta \,
              \wp(\beta(q)) \, \Phi(\gamma(q),u) \;
              N_{\gamma-\beta,\beta} \, E_\gamma~.
\end{eqnarray*}
Renaming summation indices in the last two sums ($\alpha \rightarrow
-\alpha \,$ in the first, $\beta \rightarrow \alpha \,$ in the second)
and using eqns (\ref{eq:NORM2}), (\ref{eq:CYCID}), (\ref{eq:COUPC1})
and (\ref{eq:POTF1}), we finally obtain
\begin{eqnarray*}
 \dot{L}(u) \!\!\!&-&\!\!\! [L(u),A(u)]                                 \\[3mm]
 &=&\!\! \dot{p} \,
         + \; {\textstyle {1 \over 2}} \,
              \sum_{\alpha \ssmin \Delta} \mbox{\sl g}_\alpha^2 \;
              \wp^\prime(\alpha(q)) \; H_\alpha \,
         + \, \sum_{\alpha \ssmin \Delta} i \, \mbox{\sl g}_\alpha \,
              \Phi^\prime(\alpha(q),u) \, \alpha(\dot{q} - p) \, E_\alpha    \\
 & &\!\! + \; {\textstyle {1 \over 2}} \,
              \sum_{\alpha,\gamma \ssmin \Delta}
              \wp(\alpha(q)) \; \mbox{\sl g}_\alpha
              \left( 2 \, \mbox{\sl g}_\gamma \, \gamma(F_\alpha) \, - \,
                     \mbox{\sl g}_{\gamma+\alpha} \, N_{\gamma,\alpha} \, - \,
                     \mbox{\sl g}_{\gamma-\alpha} \, N_{\gamma,-\alpha} \right)
              \Phi(\gamma(q),u) \; E_\gamma~.
\end{eqnarray*}
The vanishing of the sum of the first two terms and of the third term
are precisely the equations of motion, so the last double sum must vanish.
Thus we are led to conclude that, for the elliptic Calogero model, the Lax
equations (\ref{eq:ELAXEQ}) will be equivalent to the Hamiltonian equations
of motion (\ref{eq:EHAMEQ}) if and only if for all roots $\, \gamma \smin
\Delta$, we have
\begin{equation} \label{eq:EKOMBI}
 \sum_{\alpha \ssmin \Delta}
 \wp(\alpha(q)) \; \mbox{\sl g}_\alpha
 \left( 2 \, \mbox{\sl g}_\gamma \, \gamma(F_\alpha) \,
        - \, {\mit{\Gamma}}_{\gamma,\alpha} \,
        - \, {\mit{\Gamma}}_{\gamma,-\alpha} \right) \!~=~0~.
\end{equation}
The same calculation with $\Phi(s,u)$ replaced by $w(s)$ and $\wp(s)$ replaced
by $\, w^{\prime\prime}(s)/2 \>\! w(s)$ \linebreak leads to the conclusion
that, for the ordinary (rational or trigonometric) Calogero \mbox{models,}
the Lax equations (\ref{eq:OLAXEQ}) will be equivalent to the Hamiltonian
equations of motion (\ref{eq:OHAMEQ}) if and only if for all roots
$\, \gamma \smin \Delta$, we have
\begin{equation} \label{eq:OKOMBI}
 \sum_{\alpha \ssmin \Delta}
 {w^{\prime\prime}(\alpha(q)) \over 2\,w(\alpha(q))} \; \mbox{\sl g}_\alpha
 \left( 2 \, \mbox{\sl g}_\gamma \, \gamma(F_\alpha) \,
        - \, {\mit{\Gamma}}_{\gamma,\alpha} \,
        - \, {\mit{\Gamma}}_{\gamma,-\alpha} \right) \!~=~0~.
\end{equation}
Since these relations must hold identically in $q$, we are led to postulate
the validity of the following combinatoric identity:
\begin{equation} \label{eq:KOMBI1}
 2 \, \mbox{\sl g}_\beta \, \beta(F_\alpha)~
 =~{\mit{\Gamma}}_{\beta,\alpha} \, + \, {\mit{\Gamma}}_{\beta,-\alpha}
 \qquad \mbox{for $\, \alpha,\beta \smin \Delta$}~.
\vspace{1mm}
\end{equation}
Passing to the Poisson bracket relation (\ref{eq:ERMAT1}) in the elliptic
case, we use eqn (\ref{eq:ELAXPL}) to compute the {\em lhs}, starting out
from the canonical Poisson brackets
$$
 \{p_j,q_k\}~=~\delta_{jk}~,
$$
which give
\vspace{1mm}
$$
 \{p_j,\Phi(\alpha(q),u)\}~=~\Phi^\prime(\alpha(q),u) \; \alpha(H_j)~,
\vspace{1mm}
$$
and hence, due to the identities (\ref{eq:RTIDE1}) and (\ref{eq:ORTIMP}),
$$
 \{p,\Phi(\alpha(q),u)\}~=~\Phi^\prime(\alpha(q),u) \; H_\alpha~.
$$
This leads to the following momentum independent expression for the {\em lhs}
of eqn (\ref{eq:ERMAT1}):
\begin{equation} \label{eq:ERMAT3}
 \{L_1(u),L_2(v)\}~
 =~\sum_{\alpha \ssmin \Delta} i \, \mbox{\sl g}_\alpha \, 
   \Bigl( \Phi^\prime(\alpha(q),v) \, H_\alpha \otimes E_\alpha \, - \,
          \Phi^\prime(\alpha(q),u) \, E_\alpha \otimes H_\alpha \,\Bigr)~.
\end{equation}
To compute the {\em rhs} of eqn (\ref{eq:ERMAT1}) we observe first that it is
also momentum independent, since the only possibly momentum dependent terms
cancel:
\begin{eqnarray*}
\lefteqn{[R_{12}(u,v),p_1] \, - \, [R_{21}(v,u),p_2]}      \hspace{1cm} \\[2mm]
 &=&\!\! \sum_{\alpha \ssmin \Delta} \alpha(p) \,
         \Bigl( \Phi(\alpha(q),u-v) \, E_\alpha \otimes E_{-\alpha} \, - \,
                \Phi(\alpha(q),v-u) \, E_{-\alpha} \otimes E_\alpha \, \Bigr)
                                                                        \\[2mm]
 &=&\!\! 0~.
\end{eqnarray*}
The remaining terms are, according to eqn (\ref{eq:RTIDE1}),
\begin{eqnarray*}
\lefteqn{[R_{12}(u,v),L_1(u) - p_1] \, - \, [R_{21}(v,u),L_2(v) - p_2]}
                                                           \hspace{1cm} \\[3mm]
 &=&\!\! - \; \sum_{\alpha \ssmin \Delta} i \, \mbox{\sl g}_\alpha \,
              \left( \zeta(u-v) + \zeta(v) \right) \Phi(\alpha(q),u) \;
              E_\alpha \otimes H_\alpha                                      \\
 & &\!\! + \; \sum_{\alpha \ssmin \Delta} i \, \mbox{\sl g}_\alpha \,
              \left( \zeta(v-u) + \zeta(u) \right) \Phi(\alpha(q),v) \;
              H_\alpha \otimes E_\alpha                                      \\
 & &\!\! + \; \sum_{\alpha,\beta \ssmin \Delta}
              \Phi(\alpha(q),u) \, \Phi(\beta(q),v) \,
              \Bigl( i \, \mbox{\sl g}_\alpha \, \alpha(F_\beta) \, - \,
                     i \, \mbox{\sl g}_\beta \, \beta(F_\alpha) \Bigr) \;
              E_\alpha \otimes E_\beta                                       \\
 & &\!\! + \; \sum_{\alpha \ssmin \Delta} i \, \mbox{\sl g}_\alpha \,
              \Phi(-\alpha(q),u-v) \, \Phi(\alpha(q),u) \;
              H_\alpha \otimes E_\alpha                                      \\
 & &\!\! - \; \sum_{\alpha \ssmin \Delta} i \, \mbox{\sl g}_\alpha \,
              \Phi(-\alpha(q),v-u) \, \Phi(\alpha(q),v) \;
              E_\alpha \otimes H_\alpha                                      \\
 & &\!\! - \; \sum_{\alpha,\beta,\gamma \ssmin \Delta \atop
                    \alpha + \beta = \gamma}
              i \, \mbox{\sl g}_\gamma \,
              \Phi(-\beta(q),u-v) \, \Phi(\gamma(q),u) \, N_{-\beta,\gamma} \;
              E_\alpha \otimes E_\beta                                       \\
 & &\!\! + \; \sum_{\alpha,\beta,\gamma \ssmin \Delta \atop
                    \alpha + \beta = \gamma}
              i \, \mbox{\sl g}_\gamma \,
              \Phi(-\alpha(q),v-u) \, \Phi(\gamma(q),v) \,
              N_{-\alpha,\gamma} \; E_\alpha \otimes E_\beta~.
\end{eqnarray*}
where the last four terms have been obtained by splitting each of the two
terms containing tensor products of a root generator with the commutator of
two other root generators, say $\, [E_\alpha,E_\beta] \otimes E_\gamma \,$
or $\, E_\gamma \otimes [E_\alpha,E_\beta]$, into two contributions: one
corresponding to terms where $\, \alpha + \beta = 0$ and one corresponding
to terms where $\, \alpha + \beta \smin \Delta$; moreover, some renaming
of summation indices has been performed. Using eqns (\ref{eq:NORM2}) and
(\ref{eq:CYCID}) and rearranging terms, we get
\begin{eqnarray*}
\lefteqn{[R_{12}(u,v),L_1(u) - p_1] \, - \, [R_{21}(v,u),L_2(v) - p_2]} \\[3mm]
 &=&\!\! + \; \sum_{\alpha \ssmin \Delta} i \, \mbox{\sl g}_\alpha \,
              \Bigl( \left( \zeta(v-u) + \zeta(u) \right) \Phi(\alpha(q),v) \;
                     + \; \Phi(-\alpha(q),u-v) \, \Phi(\alpha(q),u) \Bigr) \;
              H_\alpha \otimes E_\alpha                                      \\
 & &\!\! - \; \sum_{\alpha \ssmin \Delta} i \, \mbox{\sl g}_\alpha \,
              \Bigl( \left( \zeta(u-v) + \zeta(v) \right) \Phi(\alpha(q),u) \;
                     + \; \Phi(-\alpha(q),v-u) \, \Phi(\alpha(q),v) \Bigr) \;
              E_\alpha \otimes H_\alpha                                      \\
 & &\!\! + \; \sum_{\alpha,\beta \ssmin \Delta}
              \Phi(\alpha(q),u) \, \Phi(\beta(q),v) \,
              \Bigl( i \, \mbox{\sl g}_\alpha \, \alpha(F_\beta) \, - \,
                     i \, \mbox{\sl g}_\beta \, \beta(F_\alpha) \Bigr) \;
              E_\alpha \otimes E_\beta                                       \\
 & &\!\! + \; \sum_{\alpha,\beta,\gamma \ssmin \Delta \atop
                    \alpha + \beta = \gamma}
              i \, \mbox{\sl g}_\gamma \,
              \Bigl( \Phi(-\beta(q),u-v) \, \Phi(\gamma(q),u)          \\[-6mm]
 & &\!\! \hphantom{+ \; \sum_{\alpha,\beta,\gamma \ssmin \Delta \atop
                              \alpha + \beta = \gamma}
                        i \, \mbox{\sl g}_\gamma \, \Bigl(} + \,
                     \Phi(-\alpha(q),v-u) \, \Phi(\gamma(q),v) \Bigr) \;
                     N_{\alpha,\beta} \; E_\alpha \otimes E_\beta~.
\end{eqnarray*}
Inserting the functional equations (\ref{eq:EFUNEQ5}) and (\ref{eq:EFUNEQ6}),
we are finally left with

\pagebreak

\begin{eqnarray*}
\lefteqn{[R_{12}(u,v),L_1(u) - p_1] \, - \, [R_{21}(v,u),L_2(v) - p_2]} \\[3mm]
 &=&\!\! \sum_{\alpha \ssmin \Delta} i \, \mbox{\sl g}_\alpha \,
         \Bigl( \Phi^\prime(\alpha(q),v) \, H_\alpha \otimes E_\alpha \, - \,
                \Phi^\prime(\alpha(q),u) \, E_\alpha \otimes H_\alpha \, \Bigr)
                                                                             \\
 & &\!\! + \; \sum_{\alpha,\beta \ssmin \Delta}
              \Phi(\alpha(q),u) \, \Phi(\beta(q),v) \,
              \Bigl( i \, \mbox{\sl g}_\alpha \, \alpha(F_\beta) \, - \,
                     i \, \mbox{\sl g}_\beta \, \beta(F_\alpha) \, - \,
                     i \, \mbox{\sl g}_{\alpha+\beta} \, N_{\alpha,\beta}
                     \Bigr) \; E_\alpha \otimes E_\beta~.
\end{eqnarray*}
The first term gives precisely the {\em rhs} of eqn (\ref{eq:ERMAT3}),
so the last double sum must vanish. Thus we are led to conclude that, for
the elliptic Calogero model, the Poisson bracket relation (\ref{eq:ERMAT1})
will hold if and only if the following combinatoric identity is valid:
\begin{equation} \label{eq:KOMBI2}
 \mbox{\sl g}_\alpha \, \alpha(F_\beta) \, - \,
 \mbox{\sl g}_\beta  \, \beta(F_\alpha)~
 =~{\mit{\Gamma}}_{\alpha,\beta}
 \qquad \mbox{for $\, \alpha,\beta \smin \Delta$}~.
\end{equation}
The same calculation with $\Phi(s,u)$ and $\Phi(s,v)$ both replaced by $w(s)$,
$\Phi(s,u-v)$ and $\Phi(s,v-u)$ both replaced by $\, - w^\prime(s)/w(s) \,$
and $\zeta$ replaced by zero leads to the conclusion that, for the ordinary
(rational or trigonometric) Calogero model, the Poisson bracket relation
(\ref{eq:ORMAT1}) will hold if and only if the same combinatoric identity
(\ref{eq:KOMBI2}) is valid. Moreover, this identity is easily shown to be
equivalent to the previously imposed identity (\ref{eq:KOMBI1}).

The two basic results thus obtained are not entirely independent because the
Hamiltonian is a quadratic function of the $L$-matrix whereas the $A$-matrix
is essentially just the composition of the $R$-matrix with the $L$-matrix.
More precisely, using the invariant bilinear form $(.\,,.)$ on $\mathfrak{g}$
to identify $\mathfrak{g}$ with its dual space $\mathfrak{g}^*$ and to re%
interpret $R$ as a linear mapping from $\mathfrak{g}$ to $\mathfrak{g}$
rather than as a tensor in $\, \mathfrak{g} \otimes \mathfrak{g}$,%
\footnote{More precisely, the convention used is to apply the identification
between $\mathfrak{g}$ and $\mathfrak{g}^*$ to the second factor in the tensor
product $\, \mathfrak{g} \otimes \mathfrak{g}$.} we can form the composition
$R \cdot L$ and $\, R(u,v) \cdot L(u)$, respectively; then
\begin{equation} \label{eq:ORELHL}
 H~=~{\textstyle {1 \over 2}} \, (L,L)
\end{equation}
and, due to eqn ({\ref{eq:OFUNEQ3}),
\begin{equation} \label{eq:ORELRA}
 A~=~R \cdot L
\end{equation}
for the ordinary (rational or trigonometric) Calogero model while
\begin{equation} \label{eq:ERELHL}
 H~=~{\rm Res} \vert_{u=0}^{}
     \left( {1 \over 2u} \, (L(u),L(u)) \right)
\end{equation}
and, due to eqns ({\ref{eq:EFUNEQ2}) and (\ref{eq:EFUNEQ6}),
$$
 R(u,v) \cdot L(v)~
 =~A(u) \, - \left( \zeta(u-v) + \zeta(v) \right) L(u) \, + \,
   i \, \wp(v) \, \sum_{\alpha \ssmin \Delta} \mbox{\sl g}_\alpha F_\alpha
$$
and hence
\begin{equation} \label{eq:ERELRA}
 A(u)~=~{\rm Res} \vert_{v=u}^{}
        \left( {1 \over v-u} \, R(u,v) \cdot L(v) \right) + \, \zeta(u) \, L(u)
\end{equation}
for the elliptic Calogero model, provided that the function $F$ mentioned
above satisfies the simple constraint equation
\begin{equation} \label{eq:FUNF3}
 \sum_{\alpha \ssmin \Delta} \mbox{\sl g}_\alpha F_\alpha~=~0~.
\end{equation}
Therefore, using the general fact that the Poisson bracket relations (\ref%
{eq:ORMAT1}) and (\ref{eq:ERMAT1}) imply the Lax equations (\ref{eq:OLAXEQ})
and (\ref{eq:ELAXEQ}) if one simply substitutes $R \cdot L$ for $A$ and
$\, R(u,v) \cdot L(u) \,$ for $A(u)$, respectively, shows that with the
above choices the Lax equations follow from the Poisson bracket relations.
(For the convenience of the reader, the proof of this general fact is
reproduced in Appendix 2.)

Concluding, we mention the fact that in view of the invariance of the
structure constants $N_{\alpha,\beta}$ under the action of the Weyl group,
\begin{equation}
 N_{w\alpha,w\beta}~=~N_{\alpha,\beta} \qquad
 \mbox{for all $\, w \smin W(\mathfrak{g})$}~,
\end{equation}
together with that of the coupling constants $\mbox{\sl g}_\alpha$, as
required in eqn (\ref{eq:COUPC2}), the function $F$ must be covariant
(equivariant) under the action of the Weyl group:
\begin{equation} \label{eq:FUNF4}
 F_{w\alpha}~=~w(F_\alpha) \qquad
 \mbox{for all $\, w \smin W(\mathfrak{g})$}~.
\end{equation}

\subsection{Solution of the Combinatoric Identity}

Up to now it is not clear whether and eventually for what choice of the
coupling constants $\mbox{\sl g}_\alpha$, there exists a (vector valued)
function $F$ as in eqn (\ref{eq:FUNF1}), satisfying the required combinatoric
identity (\ref{eq:KOMBI1}). In the following, we want to answer this question
in full generality, for all semisimple Lie algebras.

To begin with, observe that the problem for semisimple Lie algebras is easily
reduced to that for simple Lie algebras. In fact, when a semisimple Lie algebra
is decomposed into the direct sum of its simple ideals, its root system will
be decomposed orthogonally into irreducible root systems, and since the sum
of any two roots belonging to different subsystems is not a root, the
combinatoric identity (\ref{eq:KOMBI1}) will split into separate identities,
one for each of the simple ideals, and with an independent choice of coupling
constants for each of them.

To analyze the problem for simple Lie algebras, we remark first of all
that, due to the fact that all roots of a simple Lie algebra having the
same length constitute a single Weyl group orbit, the choice of coupling
constants is severely restricted:
\begin{itemize}
 \item $\mathfrak{g}$ is simply laced -- $A_r$ ($r\!\geq\!1$),
       $D_r$ ($r\!\geq\!4$), $E_6$, $E_7$, $E_8$: \\[1mm]
       All roots have the same length, and there is only a single coupling
       constant $\mbox{\sl g}$, which should be nonzero. As a result, the
       coupling constants drop out from eqn (\ref{eq:KOMBI1}), which reduces
       to
       \begin{equation} \label{eq:KOMBI3}
        2 \, \beta(F_\alpha)~=~N_{\beta,\alpha} \, + \, N_{\beta,-\alpha}~.
       \end{equation}
 \item $\mathfrak{g}$ is not simply laced -- $B_r$ ($r\!\geq\!2$),
       $C_r$ ($r\!\geq\!3$), $F_4$, $G_2$: \\[1mm]
       The root system splits into precisely two Weyl group orbits, and there
       are precisely two coupling constants, $\mbox{\sl g}_l$ for the long
       roots and $\mbox{\sl g}_s$ for the short roots, at least one of which
       should be nonzero. In fact, we can be somewhat more precise, requiring
       that for the simple Lie algebras $\mathfrak{so}(2r+1,\mathbb{C})$
       of the $B$-series, $\mbox{\sl g}_l$ should be nonzero while for the
       simple Lie algebras $\mathfrak{sp}(2r,\mathbb{C})$ of the $C$-series,
       $\mbox{\sl g}_s$ should be nonzero; when $\, r=2$, both should be
       nonzero. Otherwise, the Hamiltonian of the corresponding Calogero
       model (as given in eqns (\ref{eq:OHAMF1}) or (\ref{eq:EHAMF1})) would
       decouple, that is, would decompose into the sum of $r$ copies of the
       same Hamiltonian for a system with only one degree of freedom, and
       such a system is trivially completely integrable.
\end{itemize}
Note also that the function $F$, if it exists, must be unique, since otherwise
all coupling constants would have to vanish. (The argument relies on the fact
that even when there are roots of different length, the space $\mathfrak{h}_%
{\mathbb{R}}^\ast$ is already generated both by the long roots and by the short
roots alone.)

A trivial case is that of the simplest of all simple Lie algebras, namely the
unique one of rank $1$, $\mathfrak{sl}(2,\mathbb{C})$. Here, we may simply set
$\, F \equiv 0 \,$ because there is just a single positive root, so that the
{\em rhs} of the combinatoric identity (\ref{eq:KOMBI1}) vanishes identically.

More generally, a particular role is played by the simple Lie algebras
$\mathfrak{sl}(n,\mathbb{C})$ of the $A$-series. To handle this case,
we first fix some notation. The correctly normalized invariant bilinear
form on $\mathfrak{sl}(n,\mathbb{C})$ is the trace form in the defining
representation, which can in fact be extended to a non-degenerate
invariant bilinear form on $\mathfrak{gl}(n,\mathbb{C})$:
\begin{equation} \label{eq:INVBFA}
 (X,Y)~=~{\rm trace} \, (XY) \qquad
 \mbox{for $\, X,Y \smin \mathfrak{gl}(n,\mathbb{C})$}~.
\end{equation}
Letting indices $\, a,b, \ldots \,$ run from $1$ to $n$, we introduce the
standard basis of $\mathfrak{gl}(n,\mathbb{C})$ consisting of the matrices
$E_{ab}$ with $1$ at the position where the $a$-th row and the $b$-th
column meet and with $0$ everywhere else; they satisfy the multiplication
rule
\begin{equation} \label{eq:MULTRA}
 E_{ab} E_{cd}~=~\delta_{bc} E_{ad}
\end{equation}
and hence the commutation relation
\begin{equation} \label{eq:COMMRA}
 [E_{ab},E_{cd}]~=~\delta_{bc} E_{ad} \, - \, \delta_{da} E_{cb}~.
\end{equation}
The Cartan subalgebra $\mathfrak{h}$ of $\, \mathfrak{g} = \mathfrak{sl}%
(n,\mathbb{C}) \,$ will be the usual one, consisting of all traceless
diagonal matrices; this can be extended to the standard Cartan subalgebra
$\hat{\mathfrak{h}}$ of \linebreak $\hat{\mathfrak{g}} = \mathfrak{gl}%
(n,\mathbb{C})$, consisting of all diagonal matrices; then $\mathfrak{h}_%
{\mathbb{R}}$ and $\hat{\mathfrak{h}}_{\mathbb{R}}$ consist of all real
traceless diagonal and of all real diagonal matrices, respectively.
Obviously, the matrices $\, H_a \equiv E_{aa} \,$ form an orthonormal
basis of $\hat{\mathfrak{h}}_{\mathbb{R}}$ and the linear functionals
$e_a$ given by projection of a diagonal matrix $H$ to its entries,
\begin{equation} \label{eq:RSBAS1}
 e_a(H)~=~H_{aa} \qquad
 \mbox{for $\, H \smin \hat{\mathfrak{h}}_{\mathbb{R}}$}~,
\end{equation}
form the orthonormal basis of $\hat{\mathfrak{h}}_{\mathbb{R}}^\ast$ dual
to the previous one. Moreover, setting
\begin{equation} \label{eq:RSBAS2}
 e~=~e_1 + \ldots + e_n~,
\end{equation}
we see that $\mathfrak{h}_{\mathbb{R}}^\ast$ can be identified with the
orthogonal complement of $e$ in $\hat{\mathfrak{h}}_{\mathbb{R}}^\ast$
and $\, r = n-1$. Next, the root system of $\mathfrak{sl}(n,\mathbb{C})$
is given by
\begin{equation} \label{eq:ROOTSA}
 \Delta~=~\{ \, \alpha_{ab} = e_a - e_b \, / \, 1 \leq a \neq b \leq n \, \}~,
\end{equation}
with root generators $\, E_{\alpha_{ab}} \equiv E_{ab}$ $(1 \leq a \neq b
\leq n)$ and with
\begin{equation} \label{eq:CARTGA}
 H_{\alpha_{ab}}~=~E_{aa} - E_{bb}~.
\end{equation}
The commutation relations (\ref{eq:COMMRA}) then lead to the following
explicit expression for the structure constants $\, N_{\alpha_{ab},
\alpha_{cd}} \equiv N_{ab,cd} \,$:
\begin{equation} \label{eq:STRCNA}
 N_{ab,cd}~=~\delta_{bc} - \delta_{ad}~,
\end{equation}
valid for $\, 1 \leq a,b,c,d \leq n \,$ with $\, a \neq b \,$ and
$\, c \neq d$. Moreover, using the aforementioned orthonormal bases
to identify $\hat{\mathfrak{h}}_{\mathbb{R}}$ and $\hat{\mathfrak{h}}_%
{\mathbb{R}}^\ast$ with $\mathbb{R}^n$, we see that the Weyl reflection
$s_{\alpha_{ab}}$ along the root $\alpha_{ab}$ operates on a vector in
$\mathbb{R}^n$ by simply permuting the $a$-th and the $b$-th component,
so the Weyl group $W(\mathfrak{sl}(n,\mathbb{C}))$ of $\mathfrak{sl}%
(n,\mathbb{C})$ is just the permutation group in $n$ letters.

With these preliminaries out of the way, we proceed to search for a collection
of matrices $\, F_{\alpha_{ab}} \equiv F_{ab} \,$ $(1 \leq a \neq b \leq n)$
that belong to $\mathfrak{h}_\mathbb{R}$, with $\, F_{ba} = F_{ab} \,$ as
required in eqn (\ref{eq:FUNF2}), satisfying the property (\ref{eq:FUNF4})
of covariance under the Weyl group and the combinatoric identity
\begin{equation} \label{eq:KOMBI4}
 2 \left( (F_{ab})_{cc} - (F_{ab})_{dd} \right)\!~
 =~2 \, \alpha_{cd}(F_{ab})~
 =~N_{cd,ab} + N_{cd,ba}~
 =~\delta_{da} - \delta_{cb} + \delta_{db} - \delta_{ca}~,
\end{equation}
valid for $\, 1 \leq a,b,c,d \leq n \,$ with $\, a \neq b \,$ and
$\, c \neq d$, corresponding to eqn (\ref{eq:KOMBI3}). Obviously, the
{\em rhs} vanishes if the sets $\{c,d\}$ and $\{a,b\}$ coincide and also
if they are disjoint. This means that the $a$-th and $b$-th entry of the
matrix $F_{ab}$ have to be equal and also that all other entries of the
matrix $F_{ab}$ have to be equal among themselves. Thus we can write
$$
 F_{ab}~=~{1 \over 2} \, \lambda_{ab} \left( E_{aa} + E_{bb} \right) + \,
          {1 \over n} \, \tau_{ab} \, 1~,
$$
with real coefficients $\lambda_{ab}$ and $\tau_{ab}$ to be determined.
If on the other hand we choose $c$ and $d$ so that the intersection of the
sets $\{c,d\}$ and $\{a,b\}$ contains precisely one element, the {\em rhs}
is equal to $\pm 1$, and the equation is solved by putting $\, \lambda_{ab}
= - 1 \,$ whereas the other coefficient is fixed to be $\, \tau_{ab} = 1 \,$
by the condition that $F_{ab}$ should be traceless. The result is the
following
\begin{theorem}
 For the complex simple Lie algebras $\mathfrak{sl}(n,\mathbb{C})$ of the
 $A$-series, there exists a non-trivial solution to the combinatoric identity
 (\ref{eq:KOMBI1}) which is given by
 \begin{equation} \label{eq:FUNF5}
  F_{ab}~=~- \, {1 \over 2} \left( E_{aa} + E_{bb} \right) \;\!
           + \, {1 \over n} \, 1~.
 \end{equation}
 This solution also satisfies the constraint equation (\ref{eq:FUNF3}).
\end{theorem}
The question whether other complex simple Lie algebras admit a similar solution
has a negative answer:
\begin{theorem}
 Let $\mathfrak{g}$ be a complex simple Lie algebra not belonging to the
 $A$-series, i.e., not isomorphic to any of the complex simple Lie algebras
 $\mathfrak{sl}(n,\mathbb{C})$. Then the combinatoric identity (\ref{eq:KOMBI1})
 has no solution.
\end{theorem}
In the proof, we shall for the sake of simplicity use the pertinent invariant
bilinear form $(.\,,.)$ to identify the space $\mathfrak{h}_{\mathbb{R}}$ with
its dual $\mathfrak{h}_{\mathbb{R}}^\ast$ and introduce a basis $\, \{ e_1 ,
\ldots , e_n \} \,$ which is orthonormal except possibly for an overall
normalization factor; then the root system $\Delta$ will be considered
as a finite subset of $\mathbb{R}^n$ and $F$ will be a map from $\Delta$
to $\mathbb{R}^n$ required to satisfy the combinatoric identity
\begin{equation} \label{eq:KOMBI5}
 2 \, \mbox{\sl g}_\beta \, (\beta,F_\alpha)~
 =~{\mit{\Gamma}}_{\beta,\alpha} \, + \, {\mit{\Gamma}}_{\beta,-\alpha}
 \qquad \mbox{for $\, \alpha,\beta \smin \Delta$}~.
\end{equation}
As observed before, when $\mathfrak{g}$ is simply laced, this relation
reduces to
\begin{equation} \label{eq:KOMBI6}
 2 \, (\beta,F_\alpha)~=~N_{\beta,\alpha} \, + \, N_{\beta,-\alpha}
 \qquad \mbox{for $\, \alpha,\beta \smin \Delta$}~.
\end{equation}
In particular, the solution for $A_n$ given by eqn (\ref{eq:FUNF5})
is recast into the form
\begin{equation} \label{eq:FUNF6}
 F_{ab}~=~- \, {1 \over 2} \left( e_a + e_b \right) \;\!
          + \, {1 \over n} \, e \qquad
 (1 \leq a \neq b \leq n)~.
\end{equation}

We begin with the case of the complex simple Lie algebras $\mathfrak{so}%
(2n,\mathbb{C})$ of the $D$-series, with $\, n \geq 4$, which are simply laced
and for which $\Delta$ consists of the following roots:
\begin{equation} \label{eq:WURZD}
  \pm \, \alpha_{kl}~=~\pm \left( e_k - e_l \right)\!~~,~~
  \pm     \beta_{kl}~=~\pm \left( e_k + e_l \right) \qquad
  (1 \leq k < l \leq n)~.
\end{equation}
Expanding the generators $\, F_{kl} \equiv F_{\pm \alpha_{kl}} \,$ and
$\, F_{kl}^\prime \equiv F_{\pm \beta_{kl}} \,$ in their components with
respect to the orthonormal basis $\, \{ e_1 , \ldots , e_n \} \,$ of
$\mathbb{R}^n$, we now argue as follows.
\begin{itemize}
 \item When the sets $\{p,q\}$ and $\{k,l\}$ are disjoint (this possibility
       exists because we are assuming $\, n \geq 4$), none of the expressions
       $\, \pm\,\alpha_{pq} \pm \alpha_{kl}$, $\pm\,\beta_{pq} \pm \alpha_{kl}$,
       $\pm\,\alpha_{pq} \pm \beta_{kl}$, $\pm\,\beta_{pq} \pm \beta_{kl} \,$
       is a root, so using eqn (\ref{eq:KOMBI6}) with $\, \beta
       = \alpha_{pq} \,$ as well as $\, \beta = \beta_{pq} \,$ and
       $\, \alpha = \alpha_{kl} \,$ as well as $\, \alpha = \beta_{kl}$,
       we infer that both $F_{kl}$ and $F_{kl}^\prime$ must be orthogonal to
       both $\alpha_{pq}$ and $\beta_{pq}$, i.e., to $e_p$ and to $e_q$.
       In other words, the components of $F_{kl}$ and of $F_{kl}^\prime$
       along any basis vector except $e_k$ and $e_l$ must vanish, i.e.,
       both $F_{kl}$ and $F_{kl}^\prime$ must be linear combinations of
       $e_k$ and $e_l$.
 \item When the sets $\{p,q\}$ and $\{k,l\}$ are equal, then once
       again none of the expressions $\, \pm\,\alpha_{pq} \pm \alpha_{kl}$,
       $\pm\,\beta_{pq} \pm \alpha_{kl}$, $\pm\,\alpha_{pq} \pm \beta_{kl}$,
       $\pm\,\beta_{pq} \pm \beta_{kl} \,$ is a root, so using eqn
       (\ref{eq:KOMBI6}) with $\, \beta = \alpha_{pq} \,$ as well
       as $\, \beta = \beta_{pq} \,$ and $\, \alpha = \alpha_{kl} \,$
       as well as $\, \alpha = \beta_{kl}$, we infer that both $F_{kl}$
       and $F_{kl}^\prime$ must be orthogonal to both $\alpha_{kl}$ and
       $\beta_{kl}$, i.e., to $e_k$ and to $e_l$. In other words, the
       components of $F_{kl}$ and of $F_{kl}^\prime$ along the basis
       vectors $e_k$ and $e_l$ must vanish.
\end{itemize}
Thus we arrive at the conclusion that eqn (\ref{eq:KOMBI6}) forces $F$
to be identically zero, which is of course a contradiction because the
{\em rhs} of eqn (\ref{eq:KOMBI6}) is not identically zero: more specifically,
it does not vanish when we set $\, \beta = \alpha_{pq} \,$ or $\, \beta =
\beta_{pq} \,$ and $\, \alpha = \alpha_{kl} \,$ or $\, \alpha = \beta_{kl}$,
assuming that the intersection of the sets $\{p,q\}$ and $\{k,l\}$ contains
precisely one element.

It is interesting to note that the first part of the above argument fails
when $\, n = 3 \,$ and a solution indeed exists in this case, with $F_{kl}$
and $F_{kl}^\prime$ proportional to the basis vector $\, \epsilon_{klm} e_m$.
This is to be expected due to the isomorphism between $\mathfrak{so}%
(6,\mathbb{C})$ and $\mathfrak{sl}(4,\mathbb{C})$, which at the level
of root systems can be implemented by expressing the basis vectors
$\, e_1,e_2,e_3 \,$ for the root system of $D_3$ in terms of the basis
vectors $\, \tilde{e}_1,\tilde{e}_2,\tilde{e}_3,\tilde{e}_4 \,$ for the
root system of $A_3$, as well as their sum $\tilde{e}$, by setting
\begin{equation} \label{eq:WURZAD31}
 e_k~=~\tilde{e}_k + \tilde{e}_4 - {\textstyle {1 \over 2}} \, \tilde{e} \qquad
 (1 \leq k \leq 3)~,
\end{equation}
which induces an isomorphism between the corresponding root systems given by
\begin{equation} \label{eq:WURZAD32}
 \pm \, \alpha_{kl}~=~\pm \tilde{\alpha}_{kl}~~,~~
 \pm     \beta_{kl}~=~\mp \epsilon_{klm} \tilde{\alpha}_{m4} \qquad
 (1 \leq k < l \leq 3)~.
\end{equation}
Thus the solution for $A_3$ given by eqn (\ref{eq:FUNF6}) can be translated
into a solution for $D_3 \,$:
\begin{equation} \label{eq:FUNF7}
 F_{kl}~=~{\textstyle {1 \over 2}} \, \epsilon_{klm} e_m~~,~~
 F_{kl}^\prime~=~- \, {\textstyle {1 \over 2}} \, \epsilon_{klm} e_m \qquad
 (1 \leq k < l \leq 3)~.
\end{equation}

The proof of the theorem for the other complex simple Lie algebras is greatly
simplified by comparing the eventual solution for a complex simple Lie algebra
$\mathfrak{g}$ with that for a complex simple Lie algebra $\mathfrak{g}^\prime$
when the latter, say, is a subalgebra of the former of the same rank. In this
case, a Cartan subalgebra of the latter also serves as a Cartan subalgebra of
the former, so that the root system $\Delta^\prime$ of $\mathfrak{g}^\prime$
can be identified with a subset of the root system $\Delta$ of $\mathfrak{g}$.
It is then clear that the solution for $\mathfrak{g}$ must restrict to the
solution for $\mathfrak{g}^\prime$. More specifically, this means that for
$\, \alpha \smin \Delta^\prime \smsubset \Delta$, the generator $F_\alpha$
for $\mathfrak{g}$ must (possibly up to a universal normalization factor%
\footnote{If the index of the embedding of $\mathfrak{g}^\prime$ into
$\mathfrak{g}$ is not equal to $1$, some formulas will be modified
because it will appear as a normalization factor relating the invariant
bilinear forms for $\mathfrak{g}^\prime$ and for $\mathfrak{g}$.}) coincide
with the generator $F_\alpha$ for $\mathfrak{g}^\prime$. In particular, this
makes it obvious that the nonexistence of a solution for $\mathfrak{g}^\prime$
will automatically imply the nonexistence of a solution for $\mathfrak{g}$.

This argument can be immediately applied choosing $\mathfrak{g}^\prime$ to
be one of the complex simple Lie algebras $\mathfrak{so}(2n,\mathbb{C})$ of
the $D$-series, with $\, n \geq 4$, to prove the theorem for almost all
other complex simple Lie algebras:
\begin{itemize}
 \item $B_n$ ($n \geq 4$): $B_n$ contains $D_n$,
 \item $C_n$ ($n \geq 4$): $C_n$ contains $D_n$,
 \item $E_8$: $E_8$ contains $D_8$,
 \item $F_4$: $F_4$ contains $D_4$.
\end{itemize}
Therefore, all that remains is to analyze a few isolated cases, namely
$B_2 = C_2$, $G_2$, $B_3$ and $C_3$, $E_6$ and $E_7$. This can be done
by applying similar arguments to the ones already used before.
\begin{itemize}
 \item $B_2 = C_2 \,$ or $\, \mathfrak{so}(5,\mathbb{C}) \cong
       \mathfrak{sp}(4,\mathbb{C}) \,$: \\[1mm]
       The root system consists of four long roots and four short roots,
       which can be written in the form
       \vspace{3mm}
       \begin{equation} \label{eq:WURZB2}
        \begin{array}{cc}
         \mbox{long roots:}  & \pm \, e_1 \pm e_2                       \\[1mm]
         \mbox{short roots:} & \pm \, e_1~, \; \pm e_2
        \end{array}
       \end{equation}
       when $\mathfrak{g}$ is realized as $\mathfrak{so}(5,\mathbb{C})$ ($B_2$)
       or in the form
       \begin{equation} \label{eq:WURZC2}
        \begin{array}{cc}
         \mbox{long roots:}  & \pm \, 2 \>\! e_1~,~\pm 2 \>\! e_2       \\[1mm]
         \mbox{short roots:} & \pm \, e_1 \pm e_2
        \end{array}
       \end{equation}
       when $\mathfrak{g}$ is realized as $\mathfrak{sp}(4,\mathbb{C})$ ($C_2$).
       Either way, analyzing the combinatoric identity (\ref{eq:KOMBI5}) with
       $\alpha$ any long root results in a contradiction (provided we take
       into account the condition that neither of the two coupling constants
       $\mbox{\sl g}_l$ and $\mbox{\sl g}_s$ should vanish): when $\beta$ is
       a long root, neither of the two expressions $\, \beta + \alpha \,$ and
       $\, \beta - \alpha \,$ will be a root, so the {\em rhs} must vanish,
       leading to the conclusion that $F_\alpha$ must vanish, but when
       $\beta$ is a short root, then one and only one of the two expressions
       $\, \beta + \alpha \,$ and $\, \beta - \alpha \,$ will be a root, so
       the {\em rhs} cannot vanish, leading to the conclusion that $F_\alpha$
       cannot vanish. \vspace{5mm}
 \item $B_3 \,$ or $\, \mathfrak{so}(7,\mathbb{C}) \,$ and
       $\, C_3 \,$ or $\, \mathfrak{sp}(4,\mathbb{C}) \,$: \\[1mm]
       The root system for $B_3$ consists of twelve long roots and six short
       roots, which can be written in the form
       \begin{equation} \label{eq:WURZB3}
        \begin{array}{cccc}
         \mbox{long roots:}  & \pm \, e_k \pm e_l & & (1 \leq k < l \leq 3)
         \\[1mm]
         \mbox{short roots:} & \pm \, e_k         & & (1 \leq k \leq 3)
        \end{array}~,
       \end{equation}
       while the root system for $C_3$ consists of six long roots and twelve
       short roots, which can be written in the form
       \begin{equation} \label{eq:WURZC3}
        \begin{array}{cccc}
         \mbox{long roots:}  & \pm \, 2 \>\! e_k  & & (1 \leq k \leq 3)
         \\[1mm]
         \mbox{short roots:} & \pm \, e_k \pm e_l & & (1 \leq k < l \leq 3)
        \end{array}~.
       \end{equation}
       Note that both systems contain the system $D_3$ as a subsystem:
       it is generated by the roots of the form $\, \pm \, e_k \pm e_l$
       ($1 \leq k < l \leq 3$). Therefore, the values of $F$ on any root
       $\alpha$ of this form must, according to eqn (\ref{eq:FUNF7}), be
       proportional to $\, \epsilon_{klm} e_m$. But then analyzing the
       combinatoric identity (\ref{eq:KOMBI5}) with $\alpha$ any root of
       this form and $\beta$ any root of the form $\pm e_m$ (for $B_3$) or
       $\pm 2 e_m$ (for $C_3$) results in a contradiction (provided we take
       into account the condition that the coupling constants $\mbox{\sl g}_l$
       for $B_3$ and $\mbox{\sl g}_s$ for $C_3$ should not vanish): when $m$
       is equal to $k$ or $l$, then one and only one of the two expressions
       $\, \beta + \alpha \,$ and $\, \beta - \alpha \,$ will be a root, so
       the {\em rhs} cannot vanish whereas the {\em lhs} is obviously equal
       to zero; similarly, when $m$ is different from both $k$ and $l$, then
       neither of the two expressions $\, \beta + \alpha \,$ and $\, \beta
       - \alpha \,$ will be a root, so the {\em rhs} must vanish, whereas
       the {\em lhs} is typically different from zero (except when the
       other coupling constant ($\mbox{\sl g}_s$ for $B_3$ and
       $\mbox{\sl g}_l$ for $C_3$) vanishes).
 \item $G_2 \,$: \\[1mm]
       The root system consists of six long roots and six short roots,
       which can be written in the form
       \begin{equation} \label{eq:WURZG}
        \begin{array}{cccc}
         \mbox{long roots:}  &
         \pm \,  \beta_{kl}~=~\pm (2 \epsilon_{jkl} e_j - e_k - e_l) & &
         (1 \leq k < l \leq 3) \\[1mm]
         \mbox{short roots:} &
         \pm \, \alpha_{kl}~=~\pm (e_k - e_l) & &
         (1 \leq k < l \leq 3)
        \end{array}~,
       \end{equation}
       spanning the subspace orthogonal to the vector $\, e = e_1 + e_2 + e_3$.
       Note that this system is the disjoint union of two subsystems that are
       copies of the system $A_2$. \linebreak In particular, the value of $F$
       on any short root $\alpha_{kl}$ must, according to eqn (\ref{eq:FUNF6}),
       be proportional to the long root $\beta_{kl}$. But then analyzing the
       combinatoric identity (\ref{eq:KOMBI5}) with $\alpha$ any short root
       and $\beta$ the corresponding long root results in a contradiction
       (provided we take into account the condition that neither of the
       two coupling constants $\mbox{\sl g}_l$ and $\mbox{\sl g}_s$ should
       vanish because otherwise, we would really be dealing not with a $G_2$
       model but with an $A_2$ model in disguise): since neither of the two
       expressions $\, \beta_{kl} + \alpha_{kl} \,$ and $\, \beta_{kl} -
       \alpha_{kl} \,$ will be a root, the {\em rhs} must vanish, whereas
       the {\em lhs} is typically different from zero (except when the
       coupling constant $\mbox{\sl g}_l$ vanishes).
 \item $E_7 \,$: \\[1mm]
       The root system consists of $126$ roots, all of the same length,
       spanning a $7$-dimensional subspace of $\mathbb{R}^8$, which can
       be written in the form
       \begin{equation}
        \begin{array}{rcl}
         \pm \, \alpha_{kl} \!\!&=&\!\! \pm \, (e_k - e_l) \quad
         (1 \leq k < l \leq 6)                                          \\[1mm]
         \pm \, \alpha_{78} \!\!&=&\!\! \pm \, (e_7 - e_8)              \\[1mm]
         \pm \,  \beta_{kl} \!\!&=&\!\! \pm \, (e_k + e_l) \quad
         (1 \leq k < l \leq 6)                                          \\[1mm]
         \pm \, \gamma_{\pm,\dots,\pm} \!\!&=&\!\! \pm \, {1 \over 2}
         \left( e_8 - e_7 + {\displaystyle {\sum_{k=1}^6}} \,
                (-1)^{s(k)} \, e_k \right)
        \end{array}~,
       \end{equation}
       where the $s(k)$ ($k = 1,\ldots,6$) are $0$ or $1$ and such that
       $~\sum_{k=1}^6 \, s(k)~$ is odd. Applying the same arguments as
       in the $D_n$ case, we see that $F_{\alpha_{kl}}$ and $F_{\beta_{kl}}$
       \linebreak
       ($1 \leq k < l \leq 6$) must be orthogonal to $e_k$ ($k = 1,\ldots,6$)
       as well as to $\, e_7 - e_8 \,$ and can therefore be expressed in the
       form
       $$
        \begin{array}{rcl}
         F_{\alpha_{kl}} \!\!&=&\!\! f_{\alpha_{kl}} \, (e_7 + e_8)     \\[1mm]
         F_{\beta_{kl}}  \!\!&=&\!\! f_{\beta_{kl}}  \, (e_7 + e_8)
        \end{array}~.
       $$
       Obviously, these generators are also orthogonal to all roots of the
       form $\gamma_{\pm,\dots,\pm}$. But then analyzing the combinatoric
       identity (\ref{eq:KOMBI6}) with $\, \alpha = \alpha_{kl} \,$ or
       $\, \alpha = \beta_{kl} \,$ and $\beta$ an appropriate root of
       the form $\gamma_{\pm,\dots,\pm}$ results in a contradiction.
       For example, letting $\gamma_{12}$ be any root of the form
       $$
        {\textstyle {1 \over 2}}
        \left( e_8 - e_7 - (e_1 - e_2) \pm e_3 \pm e_4 \pm e_5 \pm e_6
               \right)
       $$
       with an even number of minus signs, we see that $\, \gamma_{12}
       - \alpha_{12} \,$ is not a root whereas $\, \gamma_{12}
       + \alpha_{12} \,$ is a root, so the {\em rhs} cannot vanish,
       whereas the {\em lhs} is necessarily equal to zero.
 \item $E_6 \,$: \\[1mm]
       The root system consists of $72$ roots, all of the same length,
       spanning a $6$-dimensional subspace of $\mathbb{R}^8$, which can
       be written in the form
       \begin{equation}
        \begin{array}{rcl}
         \pm \, \alpha_{kl} \!\!&=&\!\! \pm \, (e_k - e_l) ~\qquad~
         (1 \leq k < l \leq 5)                                          \\[1mm]
         \pm \,  \beta_{kl} \!\!&=&\!\! \pm \, (e_k + e_l) ~\qquad~
         (1 \leq k < l \leq 5)                                          \\[1mm]
         \pm \, \gamma_{\pm,\dots,\pm} \!\!&=&\!\! \pm \, {1 \over 2}
         \left( e_8 - e_7 - e_6 + {\displaystyle {\sum_{k=1}^5}} \,
                (-1)^{s(k)} \, e_k \right)
        \end{array}~,
       \end{equation}
       where the $s(k)$ ($k = 1,\ldots,5$) are $0$ or $1$ and such that
       $~\sum_{k=1}^5 \, s(k)~$ is even. Applying the same arguments as
       in the $D_n$ case, we see that $F_{\alpha_{kl}}$ and $F_{\beta_{kl}}$
       must be orthogonal to $e_k$ with $\, k = 1,\ldots,5 \,$ and can
       therefore be expressed in the form
       $$
        \begin{array}{rcl}
         F_{\alpha_{kl}} \!\!&=&\!\! f_{\alpha_{kl}} e_6 \, + \,
                                     g_{\alpha_{kl}} e_7 \, + \,
                                     h_{\alpha_{kl}} e_8                \\[1mm]
         F_{\beta_{kl}}  \!\!&=&\!\! f_{\beta_{kl}} e_6 \, + \,
                                     g_{\beta_{kl}} e_7 \, + \,
                                     h_{\beta_{kl}} e_8
        \end{array}~.
       $$
       To show that these generators are also orthogonal to the vector
       $\, e_8 - e_7 - e_6 \,$ and hence to all roots of the form
       $\gamma_{\pm,\dots,\pm}$, we use the combinatoric identity
       (\ref{eq:KOMBI6}) with $\, \alpha = \alpha_{kl} \,$ or
       $\, \alpha = \beta_{kl} \,$ and $\beta$ an appropriate
       root of the form $\gamma_{\pm,\dots,\pm}$. For example,
       letting $\gamma_{12}^-$ be any root of the form
       $$
        {\textstyle {1 \over 2}}
        \left( e_8 - e_7 - e_6 + (e_1 + e_2) \pm e_3 \pm e_4 \pm e_5
               \right)
       $$
       with an even number of minus signs and letting $\gamma_{12}^+$
       be any root of the form
       $$
        {\textstyle {1 \over 2}}
        \left( e_8 - e_7 - e_6 + (e_1 - e_2) \pm e_3 \pm e_4 \pm e_5
               \right)
       $$
       with an odd number of minus signs, we see that neither
       $\, \gamma_{12}^- + \alpha_{12} \,$ nor $\, \gamma_{12}^-
       - \alpha_{12} \,$ is a root and neither $\, \gamma_{12}^+
       + \beta_{12} \,$ nor $\, \gamma_{12}^+ - \beta_{12} \,$
       is a root, forcing $F_{\alpha_{12}}$ to be orthogonal to
       $\gamma_{12}^-$ and $F_{\beta_{12}}$ to be orthogonal to
       $\gamma_{12}^+$, so both must be orthogonal to the vector
       $\, e_8 - e_7 - e_6 \,$ and hence to all roots of the form
       $\gamma_{\pm,\dots,\pm}$. But then analyzing the combinatoric
       identity (\ref{eq:KOMBI6}) with $\, \alpha = \alpha_{kl} \,$ or
       $\, \alpha = \beta_{kl} \,$ and $\beta$ another appropriate root
       of the form $\gamma_{\pm,\dots,\pm}$ results in a contradiction.
       For example, letting $\gamma_{12}$ be any root of the form
       $$
        {\textstyle {1 \over 2}}
        \left( e_8 - e_7 - e_6 - (e_1 - e_2) \pm e_3 \pm e_4 \pm e_5
               \right)
       $$
       with an odd number of minus signs, we see that $\, \gamma_{12}
       - \alpha_{12} \,$ is not a root whereas $\, \gamma_{12}
       + \alpha_{12} \,$ is a root, so the {\em rhs} cannot vanish,
       whereas the {\em lhs} is necessarily equal to zero.
\end{itemize}
This completes the proof of our no-go theorem, Theorem 3.2.

To conclude this section, we briefly summarize the main results. The only
integrable Calogero models that allow a Lax formulation with a dynamical
$R$-matrix, directly in terms of simple Lie algebras, are the ones based
on the simple Lie algebras $\mathfrak{sl}(n,\mathbb{C})$ of the $A$-series.
Using the abbreviation
\begin{equation} \label{eq:POTF}
 V(t)~=~\left\{ \begin{array}{cc}
                 {1 \over 2} \, w(t)^2 & \mbox{for the ordinary (rational
                                               or trigonometric) case} \\[1mm]
                 {1 \over 2} \, \wp(t) & \mbox{for the elliptic case}
                \end{array} \right\}~,
\end{equation}
for the potential, their Hamiltonian (\ref{eq:SOHAMF1}) or (\ref{eq:SEHAMF1})
is given by
\begin{equation} \label{eq:HAMFA}
 H(q,p)~=~{\textstyle {1 \over 2}} \, \sum_{k=1}^n \, p_k^2 \, + \,
          \mbox{\sl g}^2 \! \sum_{1 \leq k \neq l \leq n} \!
          V(q_k - q_l)~,
\end{equation}
which is the Calogero Hamiltonian associated to the root system $A_{n-1}$.
Inserting the solution (\ref{eq:FUNF5}) of the combinatoric identity
(\ref{eq:KOMBI1}) reproduces the known results, both for the Lax pair
\cite{Mo,OP1,OP2,Pe} and the $R$-matrix \cite{AT,Sk}. Moreover, our
analysis shows why it is not possible to extend this technique to
other simple Lie algebras.

\section{Calogero Models for Symmetric Pairs}

\subsection{Definition of the Models, Lax Pairs and R-Matrices}

The formulation of the Calogero models associated with restricted root
systems of symmetric spaces, rather than ordinary root systems of Lie
algebras, requires only minor modifications. More specifically, we
introduce the Calogero model associated with the restricted root
system $\bar{\Delta}$ of the symmetric pair $(\mathfrak{g},\theta)$
as follows. The configuration space $Q$ of the theory is again a Weyl
chamber $C$ or Weyl alcove $A$ in the appropriate sense, that is, it is
now an open subset of $\mathfrak{a}_0$, and the phase space of the model
is the associated cotangent bundle $T^* Q = Q \times \mathfrak{a}_0^*$,
often identified with the tangent bundle $\, TQ = Q \times \mathfrak{a}_0$.
The Hamiltonian for the ordinary (rational or trigonometric) Calogero model
is then defined by
\begin{equation} \label{eq:SOHAMF1}
 H(q,p)~=~{1 \over 2} \, \Bigl( \, \sum_{j=1}^r \, p_j^2 \, + \,
          \sum_{\bar{\alpha} \ssmin \bar{\Delta}} m_{\bar{\alpha}}^{} \,
          \mbox{\sl g}_{\bar{\alpha}}^2 \, w(\bar{\alpha}(q))^2 \, \Bigr)~,
\vspace{-1mm}
\end{equation}
or equivalently
\begin{equation} \label{eq:SOHAMF2}
 H(q,p)~=~{1 \over 2} \, \Bigl( \, \sum_{j=1}^r \, p_j^2 \, + \,
          \sum_{\alpha \ssmin \tilde{\Delta}} \,
          \mbox{\sl g}_\alpha^2 \, w(\alpha(q))^2 \, \Bigr)~,
\vspace{2mm}
\end{equation}
while that for the elliptic Calogero model is defined by
\begin{equation} \label{eq:SEHAMF1}
 H(q,p)~=~{1 \over 2} \, \Bigl( \, \sum_{j=1}^r \, p_j^2 \, + \,
          \sum_{\bar{\alpha} \ssmin \bar{\Delta}} m_{\bar{\alpha}}^{} \,
          \mbox{\sl g}_{\bar{\alpha}}^2 \, \wp \>\! (\bar{\alpha}(q)) \,
          \Bigr)~,
\end{equation} 
or equivalently
\begin{equation} \label{eq:SEHAMF2}
 H(q,p)~=~{1 \over 2} \, \Bigl( \, \sum_{j=1}^r \, p_j^2 \, + \,
          \sum_{\alpha \ssmin \tilde{\Delta}} \,
          \mbox{\sl g}_\alpha^2 \, \wp \>\! (\alpha(q)) \, \Bigr)~,
\end{equation} 
where $w$ and $\wp$ are as before. The coefficients $\, \mbox{\sl g}_%
{\bar{\alpha}} = \mbox{\sl g}_\alpha$ ($\alpha \smin \tilde{\Delta}$)
are positive real coupling constants whose definition will be extended
from $\tilde{\Delta}$ to all of $\Delta$ by setting
\begin{equation} \label{eq:COUPC3}
 \mbox{\sl g}_\alpha~=~0 \qquad \mbox{for $\, \alpha \smin \Delta_0$}~.
\end{equation}
They satisfy the symmetry condition (\ref{eq:COUPC1}) and are $\theta$-%
invariant:
\begin{equation} \label{eq:COUPC4}
 \mbox{\sl g}_{-\alpha}~=~\mbox{\sl g}_\alpha~=~\mbox{\sl g}_{\theta\alpha}~.
\end{equation}
A stronger assumption that we shall always make is that they are invariant
under the action of the subgroup $W_\theta(\mathfrak{g})$ of $W(\mathfrak{g})$
associated with the symmetric pair $(\mathfrak{g},\theta)$:
\begin{equation} \label{eq:COUPC5}
 \mbox{\sl g}_{w\alpha}~=~\mbox{\sl g}_\alpha \qquad
 \mbox{for all $\, w \smin W_\theta(\mathfrak{g})$}~.
\end{equation}
For later use we introduce the following combination of coupling constants
and structure constants:
\begin{equation} \label{eq:SGAMMA}
 {\mit{\Gamma}}_{\alpha,\beta}^\theta~
 =~{\textstyle {1 \over 4}} \,
   \Bigl( \mbox{\sl g}_{\alpha+\beta} \, N_{\alpha,\beta} \, + \,
          \mbox{\sl g}_{\theta\alpha+\beta} \, N_{\theta\alpha,\beta} \, + \,
          \mbox{\sl g}_{\alpha+\theta\beta} \, N_{\alpha,\theta\beta} \, + \,
          \mbox{\sl g}_{\theta\alpha+\theta\beta} \,
                     N_{\theta\alpha,\theta\beta} \Bigr)~.
\end{equation}
Using the abbreviations (\ref{eq:ORTIMP}) as before, we can write the
Hamiltonian equations of motion as vector equations in $\mathfrak{a}_0$:
they read
\begin{equation} \label{eq:SOHAMEQ}
 \dot{q}~=~p~~~,~~~
 \dot{p}~=~- \, \sum_{\alpha \ssmin \tilde{\Delta}} \mbox{\sl g}_\alpha^2 \;
                w(\alpha(q)) \, w^\prime(\alpha(q)) \;
                (H_\alpha)_{\mathfrak{a}}
\end{equation}
for the ordinary (rational or trigonometric) Calogero model and
\begin{equation} \label{eq:SEHAMEQ}
 \dot{q}~=~p~~~,~~~
 \dot{p}~=~- \; {\textstyle {1 \over 2}} \,
                \sum_{\alpha \ssmin \tilde{\Delta}} \mbox{\sl g}_\alpha^2 \;
                \wp^\prime(\alpha(q)) \; (H_\alpha)_{\mathfrak{a}}
\end{equation}  
for the elliptic Calogero model.

The general considerations on the structure of the Lax pair and the
$R$-matrix remain unaltered, so we may pass directly to postulate a
Lax pair and an $R$-matrix; they are explicitly given by
\begin{equation} \label{eq:SOLAXPL1}
 L~=~\sum_{j=1}^r \, p_j H_j \, + \,
     \sum_{\alpha \ssmin \tilde{\Delta}} i \, \mbox{\sl g}_\alpha \,
     w(\alpha(q)) \, E_\alpha~,
\end{equation}
\begin{equation} \label{eq:SOLAXPA1}
 A~=~- \, \sum_{\alpha \ssmin \tilde{\Delta}} i \, \mbox{\sl g}_\alpha \,
     {w^{\prime\prime}(\alpha(q)) \over 2\,w(\alpha(q))} \; F_\alpha \,
     + \, \sum_{\alpha \ssmin \tilde{\Delta}} i \, \mbox{\sl g}_\alpha \,
          w^\prime(\alpha(q)) \, E_\alpha~,
\vspace{1mm}
\end{equation}
\begin{equation} \label{eq:SORMAT2}
 R~=~\sum_{\alpha \ssmin \tilde{\Delta}} w(\alpha(q)) \,
     F_\alpha \otimes E_\alpha \, + \,
     {\textstyle {1 \over 2}} \,
     \sum_{\alpha \ssmin \tilde{\Delta}}
          {w^\prime(\alpha(q)) \over w(\alpha(q))} \,
     \Bigl( E_\alpha \otimes E_{-\alpha} \, + \,
            E_{\theta\alpha} \otimes E_{-\alpha} \Bigr)~,
\vspace{2mm}
\end{equation}
for the ordinary (rational or trigonometric) Calogero model and by
\begin{equation} \label{eq:SELAXPL}
 L(u)~=~\sum_{j=1}^r \, p_j H_j \, + \,
        \sum_{\alpha \ssmin \tilde{\Delta}} i \, \mbox{\sl g}_\alpha \,
        \Phi(\alpha(q),u) \, E_\alpha~,
\end{equation}
\begin{equation} \label{eq:SELAXPA1}
 A(u)~=~- \, \sum_{\alpha \ssmin \tilde{\Delta}} i \, \mbox{\sl g}_\alpha \,
             \wp \>\! (\alpha(q)) \, F_\alpha \,
        + \, \sum_{\alpha \ssmin \tilde{\Delta}} i \, \mbox{\sl g}_\alpha \,
             \Phi^\prime(\alpha(q),u) \, E_\alpha~,
\vspace{-2mm}
\end{equation}
\begin{eqnarray} \label{eq:SERMAT2}
 R(u,v) \!\!
 &=&\!\! - \, {\textstyle {1 \over 2}} \, \sum_{j=1}^r \,
              (\zeta(u-v) + \zeta(u+v)) \, H_j \otimes H_j    \nonumber \\[1mm]
 & &\!\! - \, {\textstyle {1 \over 2}} \,
              (\zeta(u-v) - \zeta(u+v) + 2 \>\! \zeta(v)) \,
              C_{\mathfrak{k}}                                \nonumber \\[3mm]
 & &\!\! + \, \sum_{\alpha \ssmin \tilde{\Delta}}
              \Phi(\alpha(q),v) \, F_\alpha \otimes E_\alpha                 \\
 & &\!\! - \, {\textstyle {1 \over 2}} \, \sum_{\alpha \ssmin \tilde{\Delta}}
              \Bigl( \Phi(\alpha(q),u-v) \,
                     E_\alpha \otimes E_{-\alpha} \, + \,
                     \Phi(\alpha(q),-u-v) \,
                     E_{\theta\alpha} \otimes E_{-\alpha} \Bigr)
                                                        \quad \nonumber
\end{eqnarray}
for the elliptic Calogero model. Here, $C_{\mathfrak{k}}$ is the quadratic
Casimir element for $\mathfrak{k}$, i.e.,
\begin{equation}
 C_{\mathfrak{k}}~=~\sum_{j=r+1}^{r+s} H_j \otimes H_j \, + \,
                    \sum_{\alpha \ssmin \Delta_0} E_\alpha \otimes E_{-\alpha}~,
\end{equation}
and the $F_\alpha$ form a collection of generators which are now supposed to
belong to $i\mathfrak{b}_0$ and can be viewed as a (vector valued) function
\begin{equation} \label{eq:SFUNF1}
 F: \tilde{\Delta}~\longrightarrow~i\mathfrak{b}_0
\end{equation}
which we shall assume to be even and $\theta$-invariant:
\begin{equation} \label{eq:SFUNF2}
 F_{-\alpha}~=~F_\alpha~=~F_{\theta\alpha}~.
\end{equation}
We shall now once again derive a combinatoric equation which determines
$F$ completely and which, together with the functional equations discussed
in the previous section, will be sufficient to prove the equivalence between
the Hamiltonian equations of motion and the Lax equations, as well as the
validity of the Poisson bracket relations.

Basically, the proof goes as previously but it presents some additional
subtleties. As before, we concentrate on the elliptic Calogero model.
First, we use the eqns (\ref{eq:SELAXPL}) and (\ref{eq:SELAXPA1}) to
compute the following expression for the difference between the two
sides of the Lax equation (\ref{eq:ELAXEQ}):
\begin{eqnarray*}
 \dot{L}(u) \!\!\!&-&\!\!\! [L(u),A(u)]                                 \\[3mm]
 &=&\!\! \dot{p} \,
         + \, \sum_{\alpha \ssmin \tilde{\Delta}} i \, \mbox{\sl g}_\alpha \,
              \Phi^\prime(\alpha(q),u) \, \alpha(\dot{q}) \, E_\alpha \,
         - \, \sum_{\alpha \ssmin \tilde{\Delta}} i \, \mbox{\sl g}_\alpha \,
              \Phi^\prime(\alpha(q),u) \, \alpha(p) \, E_\alpha         \\[1mm]
 & &\!\! + \, \sum_{\alpha,\gamma \ssmin \tilde{\Delta}}
              \mbox{\sl g}_\alpha \, \mbox{\sl g}_\gamma \, \wp(\alpha(q)) \,
              \Phi(\gamma(q),u) \; \gamma(F_\alpha) \, E_\gamma              \\
 & &\!\! + \; {\textstyle {1 \over 2}} \,
              \sum_{\alpha \ssmin \tilde{\Delta}} \mbox{\sl g}_\alpha^2 \,
              \Bigl( \Phi(\alpha(q),u) \, \Phi^\prime(-\alpha(q),u) \, - \,
                     \Phi(-\alpha(q),u) \, \Phi^\prime(\alpha(q),u) \Bigr)
              H_\alpha                                                       \\
 & &\!\! + \; {\textstyle {1 \over 2}} \,
              \sum_{\alpha,\beta \ssmin \tilde{\Delta} \atop
                    \alpha + \beta \ssmin \Delta_0}
              \mbox{\sl g}_\alpha \, \mbox{\sl g}_\beta \,
              \Bigl( \Phi(\alpha(q),u) \, \Phi^\prime(\beta(q),u) \, - \,
                     \Phi^\prime(\alpha(q),u) \, \Phi(\beta(q),u) \Bigr) \,
              N_{\alpha,\beta} \, E_{\alpha+\beta}                           \\
 & &\!\! + \; {\textstyle {1 \over 2}} \,
              \sum_{\alpha,\beta,\gamma \ssmin \tilde{\Delta} \atop
                    \alpha + \beta = \gamma}
              \mbox{\sl g}_\alpha \, \mbox{\sl g}_\beta \,
              \Bigl( \Phi(\alpha(q),u) \, \Phi^\prime(\beta(q),u) \, - \,
                     \Phi^\prime(\alpha(q),u) \, \Phi(\beta(q),u) \Bigr) \,
              N_{\alpha,\beta} \, E_\gamma~.
\end{eqnarray*}
Here, the double sum over commutators $\, [E_\alpha,E_\beta] \,$ appearing in
the commutator between $L(u)$ and $A(u)$ has, after antisymmetrization of the
coefficients in $\alpha$ and $\beta$, been split into three contributions, one
corresponding to terms where $\, \alpha + \beta = 0$, one corresponding to
terms where $\, \alpha + \beta \smin \Delta_0 \,$ and one corresponding to
terms where $\, \alpha + \beta \smin \tilde{\Delta}$. Inserting the functional
equation (\ref{eq:EFUNEQ3}) to transform the first two of these contributions
(note that $\, \alpha + \beta \smin \Delta_0 \,$ implies $\, \alpha(q) +
\beta(q) = 0 \,$ for all $\, q \smin Q$) and the functional equation
(\ref{eq:EFUNEQ4}) to transform the third contribution (note that
$\, \alpha + \beta \smin \tilde{\Delta} \,$ implies $\, \alpha(q)
+ \beta(q) \neq 0~{\rm mod} \, \Lambda$ \linebreak for all $\, q
\smin Q$) and rearranging terms, we get
\begin{eqnarray*}
 \dot{L}(u) \!\!\!&-&\!\!\! [L(u),A(u)]                                 \\[3mm]
 &=&\!\! \dot{p} \,
         + \; {\textstyle {1 \over 2}} \,
              \sum_{\alpha \ssmin \tilde{\Delta}} \mbox{\sl g}_\alpha^2 \;
              \wp^\prime(\alpha(q)) \; H_\alpha \,
         + \, \sum_{\alpha \ssmin \tilde{\Delta}} i \, \mbox{\sl g}_\alpha \,
              \Phi^\prime(\alpha(q),u) \, \alpha(\dot{q} - p) \, E_\alpha    \\
 & &\!\! + \; {\textstyle {1 \over 4}} \,
              \sum_{\alpha,\beta \ssmin \tilde{\Delta} \atop
                    \alpha + \beta \ssmin \Delta_0}
              \mbox{\sl g}_\alpha \, \mbox{\sl g}_\beta
              \left( \wp^\prime(\alpha(q)) \, - \,
                     \wp^\prime(\beta(q)) \right)
              N_{\alpha,\beta} \, E_{\alpha+\beta}                           \\
 & &\!\! + \, \sum_{\alpha,\gamma \ssmin \tilde{\Delta}}
              \mbox{\sl g}_\alpha \, \mbox{\sl g}_\gamma \, \wp(\alpha(q)) \,
              \Phi(\gamma(q),u) \; \gamma(F_\alpha) \, E_\gamma              \\
 & &\!\! + \; {\textstyle {1 \over 2}} \,
              \sum_{\alpha,\gamma \ssmin \tilde{\Delta} \atop
                    \gamma - \alpha \ssmin \tilde{\Delta}}
              \mbox{\sl g}_\alpha \, \mbox{\sl g}_{\gamma-\alpha} \,
              \wp(\alpha(q)) \, \Phi(\gamma(q),u) \;
              N_{\alpha,\gamma-\alpha} \, E_\gamma                           \\
 & &\!\! - \; {\textstyle {1 \over 2}} \,
              \sum_{\beta,\gamma \ssmin \tilde{\Delta} \atop
                    \gamma - \beta \ssmin \tilde{\Delta}}
              \mbox{\sl g}_{\gamma-\beta} \, \mbox{\sl g}_\beta \,
              \wp(\beta(q)) \, \Phi(\gamma(q),u) \;
              N_{\gamma-\beta,\beta} \, E_\gamma~.
\end{eqnarray*}
Renaming summation indices in the last two sums ($\alpha \rightarrow
-\alpha \,$ in the first, $\beta \rightarrow \alpha \,$ in the second)
and using eqns (\ref{eq:NORM2}), (\ref{eq:CYCID}), (\ref{eq:COUPC1})
and (\ref{eq:POTF1}), together with
\begin{eqnarray*}
 \sum_{\alpha \ssmin \tilde{\Delta}} \mbox{\sl g}_\alpha^2 \;
 \wp^\prime(\alpha(q)) \; H_\alpha \!\!
 &=&\!\! {\textstyle {1 \over 2}} \,
         \sum_{\alpha \ssmin \tilde{\Delta}} \mbox{\sl g}_\alpha^2 \,
         \Bigl( \wp^\prime(\alpha(q)) \; H_\alpha \, + \,
                \wp^\prime((\theta\alpha)(q)) \; H_{\theta\alpha} \Bigr)     \\
 &=&\!\! {\textstyle {1 \over 2}} \,
         \sum_{\alpha \ssmin \tilde{\Delta}} \mbox{\sl g}_\alpha^2 \;
         \wp^\prime(\alpha(q)) \left( H_\alpha - \theta H_\alpha \right)\!~
  =~     \sum_{\alpha \ssmin \tilde{\Delta}} \mbox{\sl g}_\alpha^2 \;
         \wp^\prime(\alpha(q)) \; (H_\alpha)_{\mathfrak{a}}~,
\end{eqnarray*}
and
\begin{eqnarray*}
\lefteqn{\sum_{\alpha,\beta \ssmin \tilde{\Delta} \atop
               \alpha + \beta \ssmin \Delta_0}
         \mbox{\sl g}_\alpha \, \mbox{\sl g}_\beta
         \left( \wp^\prime(\alpha(q)) \, - \, \wp^\prime(\beta(q)) \right)
         N_{\alpha,\beta} \, E_{\alpha+\beta}}            \hspace{1cm} \\[-3mm]
 &=&\!\! {\textstyle {1 \over 2}} \,
         \sum_{\alpha,\beta \ssmin \tilde{\Delta} \atop
               \alpha + \beta \ssmin \Delta_0}
         \mbox{\sl g}_\alpha \, \mbox{\sl g}_\beta \,
         \Bigl( \left( \wp^\prime(\alpha(q)) \, - \,
                       \wp^\prime(\beta(q)) \right)
                N_{\alpha,\beta} \, E_{\alpha+\beta}                   \\[-8mm]
 & &\!\! \phantom{{\textstyle {1 \over 2}} \,
                  \sum_{\alpha,\beta \ssmin \tilde{\Delta} \atop
                        \alpha + \beta \ssmin \Delta_0}
                  \mbox{\sl g}_\alpha \, \mbox{\sl g}_\beta \,
                  \Bigl(}
              + \left( \wp^\prime((\theta\alpha)(q)) \, - \,
                       \wp^\prime((\theta\beta)(q)) \right)
                N_{\theta\alpha,\theta\beta} \, E_{\theta\alpha+\theta\beta}
         \Bigr)                                                        \\[-4mm]
 &=&\!\! {\textstyle {1 \over 2}} \,
         \sum_{\alpha,\beta \ssmin \tilde{\Delta} \atop
               \alpha + \beta \ssmin \Delta_0}
         \mbox{\sl g}_\alpha \, \mbox{\sl g}_\beta \,
         \left( \wp^\prime(\alpha(q)) \, - \, \wp^\prime(\beta(q)) \right)
         N_{\alpha,\beta}
         \left( E_{\alpha+\beta} - \theta E_{\alpha+\beta} \right)           \\
 &=&\!\! 0~,
\end{eqnarray*}
we finally obtain

\pagebreak

\begin{eqnarray*}
 \dot{L}(u) \!\!\!&-&\!\!\! [L(u),A(u)]                                 \\[3mm]
 &=&\!\! \dot{p} \,
         + \; {\textstyle {1 \over 2}} \,
              \sum_{\alpha \ssmin \tilde{\Delta}} \mbox{\sl g}_\alpha^2 \;
              \wp^\prime(\alpha(q)) \; (H_\alpha)_{\mathfrak{a}} \,
         + \, \sum_{\alpha \ssmin \tilde{\Delta}} i \, \mbox{\sl g}_\alpha \,
              \Phi^\prime(\alpha(q),u) \, \alpha(\dot{q} - p) \, E_\alpha    \\
 & &\!\! + \; {\textstyle {1 \over 2}} \,
              \sum_{\alpha,\gamma \ssmin \tilde{\Delta}}
              \wp(\alpha(q)) \; \mbox{\sl g}_\alpha
              \left( 2 \, \mbox{\sl g}_\gamma \, \gamma(F_\alpha) \, - \,
                     \mbox{\sl g}_{\gamma+\alpha} \, N_{\gamma,\alpha} \, - \,
                     \mbox{\sl g}_{\gamma-\alpha} \, N_{\gamma,-\alpha} \right)
              \Phi(\gamma(q),u) \; E_\gamma~.
\end{eqnarray*}
The vanishing of the sum of the first two terms and of the third term
are precisely the equations of motion, so the last double sum must vanish.
Thus we are led to conclude that, for the elliptic Calogero model, the Lax
equations (\ref{eq:ELAXEQ}) will be equivalent to the Hamiltonian equations
of motion (\ref{eq:SEHAMEQ}) if and only if for all roots $\, \gamma \smin
\tilde{\Delta}$, we have
\begin{equation} \label{eq:SEKOMBI}
 \sum_{\alpha \ssmin \tilde{\Delta}}
 \wp(\alpha(q)) \; \mbox{\sl g}_\alpha
 \left( 2 \, \mbox{\sl g}_\gamma \, \gamma(F_\alpha) \,
        - \, {\mit{\Gamma}}_{\gamma,\alpha} \,
        - \, {\mit{\Gamma}}_{\gamma,-\alpha} \right) \!~=~0~.
\end{equation}
The same calculation with $\Phi(s,u)$ replaced by $w(s)$ and $\wp(s)$ replaced
by $\, w^{\prime\prime}(s)/2 \>\! w(s)$ \linebreak leads to the conclusion
that, for the ordinary (rational or trigonometric) Calogero models, the Lax
equations (\ref{eq:OLAXEQ}) are equivalent to the Hamiltonian equations of
motion (\ref{eq:OHAMEQ}) if and only if for all roots $\, \gamma \smin
\tilde{\Delta}$, we have
\begin{equation} \label{eq:SOKOMBI}
 \sum_{\alpha \ssmin \tilde{\Delta}}
 {w^{\prime\prime}(\alpha(q)) \over 2\,w(\alpha(q))} \; \mbox{\sl g}_\alpha
 \left( 2 \, \mbox{\sl g}_\gamma \, \gamma(F_\alpha) \,
        - \, {\mit{\Gamma}}_{\gamma,\alpha} \,
        - \, {\mit{\Gamma}}_{\gamma,-\alpha} \right) \!~=~0~.
\end{equation}
Since these relations must hold identically in $q$ but the coefficient
functions are invariant under the substitution $\, \alpha \rightarrow
\theta\alpha$, we are led to postulate the validity of the following
combinatoric identity:
\begin{equation} \label{eq:SKOMBI1}
 2 \, \mbox{\sl g}_\beta \, \beta(F_\alpha)~
 =~{\mit{\Gamma}}_{\beta,\alpha}^\theta \, + \,
   {\mit{\Gamma}}_{\beta,-\alpha}^\theta
 \qquad \mbox{for $\, \alpha,\beta \smin \tilde{\Delta}$}~.
\end{equation}
Passing to the Poisson bracket relation (\ref{eq:ERMAT1}) in the elliptic
case, we use eqn (\ref{eq:SELAXPL}) to compute the {\em lhs}, starting out
from the same canonical Poisson brackets as before but now using the
identities (\ref{eq:RTIDE2}) and (\ref{eq:ORTIMP}) to conclude that
$$
 \{p,\Phi(\alpha(q),u)\}~
 =~\Phi^\prime(\alpha(q),u) \; (H_\alpha)_{\mathfrak{a}}~,
$$
leading to the following momentum independent expression for the {\em lhs}
of eqn (\ref{eq:ERMAT1}):
\begin{equation} \label{eq:SERMAT3}
 \{L_1(u),L_2(v)\}~
 =~\sum_{\alpha \ssmin \tilde{\Delta}} i \, \mbox{\sl g}_\alpha \,
   \Bigl( \Phi^\prime(\alpha(q),v) \,
          (H_\alpha)_{\mathfrak{a}} \otimes E_\alpha \, - \,
          \Phi^\prime(\alpha(q),u) \,
          E_\alpha \otimes (H_\alpha)_{\mathfrak{a}} \,\Bigr)~.
\end{equation}
To compute the {\em rhs} of eqn (\ref{eq:ERMAT1}) we observe first that it
is also momentum independent, since once again, the only possibly momentum
dependent terms cancel:

\pagebreak

\begin{eqnarray*}
\lefteqn{[R_{12}(u,v),p_1] \, - \, [R_{21}(v,u),p_2]}      \hspace{1cm} \\[2mm]
 & &\!\! {\textstyle {1 \over 2}} \,
         \sum_{\alpha \ssmin \tilde{\Delta}} \alpha(p) \,
         \Bigl( + \, \Phi(\alpha(q),u-v) \,
                     E_\alpha \otimes E_{-\alpha} \,
                - \, \Phi(\alpha(q),-u-v) \,
                     E_{\theta\alpha} \otimes E_{-\alpha}              \\[-6mm]
 & &\!\! \phantom{{\textstyle {1 \over 2}} \,
                  \sum_{\alpha \ssmin \tilde{\Delta}} \alpha(p) \, \Bigl(}
                - \, \Phi(\alpha(q),v-u) \,
                     E_{-\alpha} \otimes E_\alpha \,
                + \, \Phi(\alpha(q),-v-u) \,
                     E_{-\alpha} \otimes E_{\theta\alpha} \, \Bigr)    \\[-2mm]
 &=&\!\! 0~.
\end{eqnarray*}
The remaining terms are, according to eqn (\ref{eq:RTIDE2}),
\begin{eqnarray*}
\lefteqn{[R_{12}(u,v),L_1(u) - p_1] \, - \, [R_{21}(v,u),L_2(v) - p_2]}
                                                           \hspace{1cm} \\[3mm]
 &=&\!\! - \; {\textstyle {1 \over 2}} \,
              \sum_{\alpha \ssmin \tilde{\Delta}} i \, \mbox{\sl g}_\alpha \,
              \left( \zeta(u-v) + \zeta(u+v) \right) \Phi(\alpha(q),u) \;
              E_\alpha \otimes (H_\alpha)_{\mathfrak{a}}                     \\
 & &\!\! + \; {\textstyle {1 \over 2}} \,
              \sum_{\alpha \ssmin \tilde{\Delta}} i \, \mbox{\sl g}_\alpha \,
              \left( \zeta(v-u) + \zeta(v+u) \right) \Phi(\alpha(q),v) \;
              (H_\alpha)_{\mathfrak{a}} \otimes E_\alpha                     \\
 & &\!\! - \; {\textstyle {1 \over 2}} \,
              \sum_{\alpha \ssmin \tilde{\Delta}} i \, \mbox{\sl g}_\alpha \,
              \left( \zeta(u-v) - \zeta(u+v) + 2 \>\! \zeta(v) \right)
              \Phi(\alpha(q),u) \;
              E_\alpha \otimes (H_\alpha)_{\mathfrak{b}}                     \\
 & &\!\! + \; {\textstyle {1 \over 2}} \,
              \sum_{\alpha \ssmin \tilde{\Delta}} i \, \mbox{\sl g}_\alpha \,
              \left( \zeta(v-u) - \zeta(v+u) + 2 \>\! \zeta(u) \right)
              \Phi(\alpha(q),v) \;
              (H_\alpha)_{\mathfrak{b}} \otimes E_\alpha                     \\
 & &\!\! - \; {\textstyle {1 \over 2}} \,
              \sum_{\beta \ssmin \Delta_0 \atop
                    \gamma \ssmin \tilde{\Delta}} i \, \mbox{\sl g}_\gamma \,
              \left( \zeta(u-v) - \zeta(u+v) + 2 \>\! \zeta(v) \right)
              \Phi(\gamma(q),u) \, N_{-\beta,\gamma} \;
              E_{-\beta+\gamma} \otimes E_\beta                              \\
 & &\!\! + \; {\textstyle {1 \over 2}} \,
              \sum_{\alpha \ssmin \Delta_0 \atop
                    \gamma \ssmin \tilde{\Delta}} i \, \mbox{\sl g}_\gamma \,
              \left( \zeta(v-u) - \zeta(v+u) + 2 \>\! \zeta(u) \right)
              \Phi(\gamma(q),v) \, N_{-\alpha,\gamma} \;
              E_\alpha \otimes E_{-\alpha+\gamma}                            \\
 & &\!\! + \; \sum_{\alpha,\beta \ssmin \tilde{\Delta}}
              \Phi(\alpha(q),u) \, \Phi(\beta(q),v) \,
              \Bigl( i \, \mbox{\sl g}_\alpha \, \alpha(F_\beta) \, - \,
                     i \, \mbox{\sl g}_\beta \, \beta(F_\alpha) \Bigr) \;
              E_\alpha \otimes E_\beta                                       \\
 & &\!\! + \; {\textstyle {1 \over 2}} \,
              \sum_{\alpha \ssmin \tilde{\Delta}} i \, \mbox{\sl g}_\alpha \,
              \Phi(-\alpha(q),u-v) \, \Phi(\alpha(q),u) \;
              H_\alpha \otimes E_\alpha                                      \\
 & &\!\! - \; {\textstyle {1 \over 2}} \,
              \sum_{\alpha \ssmin \tilde{\Delta}} i \, \mbox{\sl g}_\alpha \,
              \Phi(-\alpha(q),-u-v) \, \Phi(\alpha(q),-u) \;
              \theta H_\alpha \otimes E_\alpha                               \\
 & &\!\! - \; {\textstyle {1 \over 2}} \,
              \sum_{\alpha \ssmin \tilde{\Delta}} i \, \mbox{\sl g}_\alpha \,
              \Phi(-\alpha(q),v-u) \, \Phi(\alpha(q),v) \;
              E_\alpha \otimes H_\alpha                                      \\
 & &\!\! + \; {\textstyle {1 \over 2}} \,
              \sum_{\alpha \ssmin \tilde{\Delta}} i \, \mbox{\sl g}_\alpha \,
              \Phi(-\alpha(q),-v-u) \, \Phi(\alpha(q),-v) \;
              E_\alpha \otimes \theta H_\alpha                               \\
 & &\!\! - \; {\textstyle {1 \over 2}} \,
              \sum_{\beta,\gamma \ssmin \tilde{\Delta} \atop
                    \beta - \gamma \ssmin \Delta_0}
              i \, \mbox{\sl g}_\gamma \,
              \Phi(-\beta(q),u-v) \, \Phi(\gamma(q),u) \,
              N_{-\beta,\gamma} \;
              E_{-\beta+\gamma} \otimes E_\beta                              \\
 & &\!\! + \; {\textstyle {1 \over 2}} \,
              \sum_{\beta,\gamma \ssmin \tilde{\Delta} \atop
                    \beta - \gamma \ssmin \Delta_0}
              i \, \mbox{\sl g}_\gamma \,
              \Phi(-\beta(q),-u-v) \, \Phi(\gamma(q),-u) \,
              N_{-\beta,\gamma} \;
              \theta E_{-\beta+\gamma} \otimes E_\beta                       \\
 & &\!\! + \; {\textstyle {1 \over 2}} \,
              \sum_{\alpha,\gamma \ssmin \tilde{\Delta} \atop
                    \alpha - \gamma \ssmin \Delta_0}
              i \, \mbox{\sl g}_\gamma \,
              \Phi(-\alpha(q),v-u) \, \Phi(\gamma(q),v) \,
              N_{-\alpha,\gamma} \;
              E_\alpha \otimes E_{-\alpha+\gamma}                            \\
 & &\!\! - \; {\textstyle {1 \over 2}} \,
              \sum_{\alpha,\gamma \ssmin \tilde{\Delta} \atop
                    \alpha - \gamma \ssmin \Delta_0}
              i \, \mbox{\sl g}_\gamma \,
              \Phi(-\alpha(q),-v-u) \, \Phi(\gamma(q),-v) \,
              N_{-\alpha,\gamma} \;
              E_\alpha \otimes \theta E_{-\alpha+\gamma}                     \\
 & &\!\! - \; {\textstyle {1 \over 2}} \,
              \sum_{\beta,\gamma \ssmin \tilde{\Delta} \atop
                    \beta - \gamma \ssmin \tilde{\Delta}}
              i \, \mbox{\sl g}_\gamma \,
              \Phi(-\beta(q),u-v) \, \Phi(\gamma(q),u) \,
              N_{-\beta,\gamma} \;
              E_{-\beta+\gamma} \otimes E_\beta                              \\
 & &\!\! + \; {\textstyle {1 \over 2}} \,
              \sum_{\beta,\gamma \ssmin \tilde{\Delta} \atop
                    \beta - \gamma \ssmin \tilde{\Delta}}
              i \, \mbox{\sl g}_\gamma \,
              \Phi(-\beta(q),-u-v) \, \Phi(\gamma(q),-u) \,
              N_{-\beta,\gamma} \;
              \theta E_{-\beta+\gamma} \otimes E_\beta                       \\
 & &\!\! + \; {\textstyle {1 \over 2}} \,
              \sum_{\alpha,\gamma \ssmin \tilde{\Delta} \atop
                    \alpha - \gamma \ssmin \tilde{\Delta}}
              i \, \mbox{\sl g}_\gamma \,
              \Phi(-\alpha(q),v-u) \, \Phi(\gamma(q),v) \,
              N_{-\alpha,\gamma} \;
              E_\alpha \otimes E_{-\alpha+\gamma}                            \\
 & &\!\! - \; {\textstyle {1 \over 2}} \,
              \sum_{\alpha,\gamma \ssmin \tilde{\Delta} \atop
                    \alpha - \gamma \ssmin \tilde{\Delta}}
              i \, \mbox{\sl g}_\gamma \,
              \Phi(-\alpha(q),-v-u) \, \Phi(\gamma(q),-v) \,
              N_{-\alpha,\gamma} \;
              E_\alpha \otimes \theta E_{-\alpha+\gamma}~,
\end{eqnarray*}
where the last twelve terms have been obtained by splitting each of the four
terms containing tensor products of a root generator with the commutator of
two other root generators, say $\, [E_\alpha,E_\beta] \otimes E_\gamma \,$
or $\, E_\gamma \otimes [E_\alpha,E_\beta]$, into three contributions: one
corresponding to terms where $\, \alpha + \beta = 0$, one corresponding to
terms where $\, \alpha + \beta \smin \Delta_0 \,$ and one corresponding to
terms where $\, \alpha + \beta \smin \tilde{\Delta}$; moreover, some re%
naming of summation indices has been performed and relations of the form
$$
 \Phi((\theta\gamma)(q),w)~=~\Phi(-\gamma(q),w)~=~- \, \Phi(\gamma(q),-w)
$$
have been employed. Using eqns (\ref{eq:NORM2}) and (\ref{eq:CYCID}) and
rearranging terms, we get
\vspace{2mm}
\begin{eqnarray*}
\lefteqn{[R_{12}(u,v),L_1(u) - p_1] \, - \, [R_{21}(v,u),L_2(v) - p_2]}
                                                           \hspace{1cm} \\[3mm]
 &=&\!\! + \; {\textstyle {1 \over 2}} \,
              \sum_{\alpha \ssmin \tilde{\Delta}} i \, \mbox{\sl g}_\alpha \,
              \Bigl( \left( \zeta(v-u) + \zeta(v+u) \right)
                     \Phi(\alpha(q),v)                                 \\[-6mm]
 & &\!\! \phantom{+ \; {\textstyle {1 \over 2}} \,
                       \sum_{\alpha \ssmin \tilde{\Delta}}
                       i \, \mbox{\sl g}_\alpha \, \Bigl(}
                     + \, \Phi(-\alpha(q),u-v) \, \Phi(\alpha(q),u)    \\[-6mm]
 & &\!\! \phantom{+ \; {\textstyle {1 \over 2}} \,
                       \sum_{\alpha \ssmin \tilde{\Delta}}
                       i \, \mbox{\sl g}_\alpha \, \Bigl(}
                     + \, \Phi(-\alpha(q),-u-v) \, \Phi(\alpha(q),-u) \Bigr) \;
              (H_\alpha)_{\mathfrak{a}} \otimes E_\alpha               \\[-3mm]
 & &\!\! - \; {\textstyle {1 \over 2}} \,
              \sum_{\alpha \ssmin \tilde{\Delta}} i \, \mbox{\sl g}_\alpha \,
              \Bigl( \left( \zeta(u-v) + \zeta(u+v) \right)
                     \Phi(\alpha(q),u)                                 \\[-6mm]
 & &\!\! \phantom{+ \; {\textstyle {1 \over 2}} \,
                       \sum_{\alpha \ssmin \tilde{\Delta}}
                       i \, \mbox{\sl g}_\alpha \, \Bigl(}
                     + \, \Phi(-\alpha(q),v-u) \, \Phi(\alpha(q),v)    \\[-6mm]
 & &\!\! \phantom{+ \; {\textstyle {1 \over 2}} \,
                       \sum_{\alpha \ssmin \tilde{\Delta}}
                       i \, \mbox{\sl g}_\alpha \, \Bigl(}
                     + \, \Phi(-\alpha(q),-v-u) \, \Phi(\alpha(q),-v) \Bigr) \;
              E_\alpha \otimes (H_\alpha)_{\mathfrak{a}}               \\[-3mm]
 & &\!\! + \; {\textstyle {1 \over 2}} \,
              \sum_{\alpha \ssmin \tilde{\Delta}} i \, \mbox{\sl g}_\alpha \,
              \Bigl( \left( \zeta(v-u) - \zeta(v+u) + 2 \>\! \zeta(u) \right)
                     \Phi(\alpha(q),v)                                 \\[-6mm]
 & &\!\! \phantom{+ \; {\textstyle {1 \over 2}} \,
                       \sum_{\alpha \ssmin \tilde{\Delta}}
                       i \, \mbox{\sl g}_\alpha \, \Bigl(}
                     + \, \Phi(-\alpha(q),u-v) \, \Phi(\alpha(q),u)    \\[-6mm]
 & &\!\! \phantom{+ \; {\textstyle {1 \over 2}} \,
                       \sum_{\alpha \ssmin \tilde{\Delta}}
                       i \, \mbox{\sl g}_\alpha \, \Bigl(}
                     - \, \Phi(-\alpha(q),-u-v) \, \Phi(\alpha(q),-u) \Bigr) \;
              (H_\alpha)_{\mathfrak{b}} \otimes E_\alpha               \\[-3mm]
 & &\!\! - \; {\textstyle {1 \over 2}} \,
              \sum_{\alpha \ssmin \tilde{\Delta}} i \, \mbox{\sl g}_\alpha \,
              \Bigl( \left( \zeta(u-v) - \zeta(u+v) + 2 \>\! \zeta(v) \right)
                     \Phi(\alpha(q),u)                                 \\[-6mm]
 & &\!\! \phantom{+ \; {\textstyle {1 \over 2}} \,
                       \sum_{\alpha \ssmin \tilde{\Delta}}
                       i \, \mbox{\sl g}_\alpha \, \Bigl(}
                     + \, \Phi(-\alpha(q),v-u) \, \Phi(\alpha(q),v)    \\[-6mm]
 & &\!\! \phantom{+ \; {\textstyle {1 \over 2}} \,
                       \sum_{\alpha \ssmin \tilde{\Delta}}
                       i \, \mbox{\sl g}_\alpha \, \Bigl(}
                     - \, \Phi(-\alpha(q),-v-u) \, \Phi(\alpha(q),-v) \Bigr) \;
              E_\alpha \otimes (H_\alpha)_{\mathfrak{b}}               \\[-3mm]
 & &\!\! + \; {\textstyle {1 \over 2}} \,
              \sum_{\alpha \ssmin \tilde{\Delta}, \beta \ssmin \Delta_0 \atop
                    \gamma = \alpha + \beta \ssmin \tilde{\Delta}}
              i \, \mbox{\sl g}_\gamma \,
              \Bigl( \left( \zeta(u-v) - \zeta(u+v) + 2 \>\! \zeta(v) \right)
                     \Phi(\gamma(q),u)                                 \\[-8mm]
 & &\!\! \phantom{+ \; {\textstyle {1 \over 2}} \,
                       \sum_{\alpha \ssmin \tilde{\Delta},
                             \beta \ssmin \Delta_0 \atop
                             \gamma = \alpha + \beta \ssmin \tilde{\Delta}}
                       i \, \mbox{\sl g}_\gamma \, \Bigl(}
                     + \, \Phi(-\alpha(q),v-u) \, \Phi(\gamma(q),v)    \\[-8mm]
 & &\!\! \phantom{+ \; {\textstyle {1 \over 2}} \,
                       \sum_{\alpha \ssmin \tilde{\Delta},
                             \beta \ssmin \Delta_0 \atop
                             \gamma = \alpha + \beta \ssmin \tilde{\Delta}}
                       i \, \mbox{\sl g}_\gamma \, \Bigl(}
                     - \, \Phi(-\alpha(q),-v-u) \, \Phi(\gamma(q),-v) \Bigr) \;
              N_{\alpha,\beta} \; E_\alpha \otimes E_\beta             \\[-5mm]
 & &\!\! + \; {\textstyle {1 \over 2}} \,
              \sum_{\alpha \ssmin \Delta_0, \beta \ssmin \tilde{\Delta} \atop
                    \gamma = \alpha + \beta \ssmin \tilde{\Delta}}
              i \, \mbox{\sl g}_\gamma \,
              \Bigl( \left( \zeta(v-u) - \zeta(v+u) + 2 \>\! \zeta(u) \right)
                     \Phi(\gamma(q),v)                                 \\[-8mm]
 & &\!\! \phantom{+ \; {\textstyle {1 \over 2}} \,
                       \sum_{\alpha \ssmin \Delta_0,
                             \beta \ssmin \tilde{\Delta} \atop
                             \gamma = \alpha + \beta \ssmin \tilde{\Delta}}
                       i \, \mbox{\sl g}_\gamma \, \Bigl(}
                     + \, \Phi(-\beta(q),u-v) \, \Phi(\gamma(q),u)     \\[-8mm]
 & &\!\! \phantom{+ \; {\textstyle {1 \over 2}} \,
                       \sum_{\alpha \ssmin \Delta_0,
                             \beta \ssmin \tilde{\Delta} \atop
                             \gamma = \alpha + \beta \ssmin \tilde{\Delta}}
                       i \, \mbox{\sl g}_\gamma \, \Bigl(}
                     - \, \Phi(-\beta(q),-u-v) \, \Phi(\gamma(q),-u) \Bigr) \;
              N_{\alpha,\beta} \; E_\alpha \otimes E_\beta             \\[-5mm]
 & &\!\! + \; \sum_{\alpha,\beta \ssmin \tilde{\Delta}}
              \Phi(\alpha(q),u) \, \Phi(\beta(q),v) \,
              \Bigl( i \, \mbox{\sl g}_\alpha \, \alpha(F_\beta) \, - \,
                     i \, \mbox{\sl g}_\beta \, \beta(F_\alpha) \Bigr) \;
              E_\alpha \otimes E_\beta                                       \\
 & &\!\! + \; {\textstyle {1 \over 2}} \,
              \sum_{\alpha,\beta,\gamma \ssmin \tilde{\Delta} \atop
                    \alpha + \beta = \gamma}
              i \, \mbox{\sl g}_\gamma \,
              \Bigl( \Phi(-\beta(q),u-v) \, \Phi(\gamma(q),u)          \\[-7mm]
 & &\!\! \phantom{+ \; {\textstyle {1 \over 2}} \,
                       \sum_{\alpha,\beta,\gamma \ssmin \tilde{\Delta} \atop
                             \alpha + \beta = \gamma}
                       i \, \mbox{\sl g}_\gamma \, \Bigl(}
                     + \, \Phi(-\alpha(q),v-u) \, \Phi(\gamma(q),v) \Bigr) \;
              N_{\alpha,\beta} \; E_\alpha \otimes E_\beta             \\[-4mm]
 & &\!\! - \; {\textstyle {1 \over 2}} \,
              \sum_{\alpha,\beta,\gamma \ssmin \tilde{\Delta} \atop
                    \alpha + \beta = \gamma}
              i \, \mbox{\sl g}_\gamma \,
              \Bigl( \Phi(-\beta(q),-u-v) \, \Phi(\gamma(q),-u)        \\[-7mm]
 & &\!\! \phantom{+ \; {\textstyle {1 \over 2}} \,
                       \sum_{\alpha,\beta,\gamma \ssmin \tilde{\Delta} \atop
                             \alpha + \beta = \gamma}
                       i \, \mbox{\sl g}_\gamma \, \Bigl(}
                     + \, \Phi(-\alpha(q),v+u) \, \Phi(\gamma(q),v) \Bigr) \;
              N_{\alpha,\beta} \; \theta E_\alpha \otimes E_\beta~.
\end{eqnarray*}
Inserting the functional equations (\ref{eq:EFUNEQ5}) and (\ref{eq:EFUNEQ6}),
we see that the terms proportional to $\, (H_\alpha)_{\mathfrak{b}} \otimes
E_\alpha \,$ and to $\, E_\alpha \otimes (H_\alpha)_{\mathfrak{b}} \,$ with
$\, \alpha \smin \tilde{\Delta}$, as well as the terms proportional to
$\, E_\alpha \otimes E_\beta \,$ with $\, \alpha \smin \tilde{\Delta} \,$
and $\, \beta \smin \Delta_0 \,$ or with $\, \alpha \smin \Delta_0 \,$
and $\, \beta \smin \tilde{\Delta} \,$ cancel, and using the convention
(\ref{eq:COUPC3}), we are finally left with
\begin{eqnarray*}
\lefteqn{[R_{12}(u,v),L_1(u) - p_1] \, - \, [R_{21}(v,u),L_2(v) - p_2]}
                                                           \hspace{1cm} \\[3mm]
 &=&\!\! \sum_{\alpha \ssmin \tilde{\Delta}} i \, \mbox{\sl g}_\alpha \,
         \Bigl( \Phi^\prime(\alpha(q),v) \,
                (H_\alpha)_{\mathfrak{a}} \otimes E_\alpha \, - \,
                \Phi^\prime(\alpha(q),u) \,
                E_\alpha \otimes (H_\alpha)_{\mathfrak{a}} \,\Bigr)          \\
 & &\!\! + \; \sum_{\alpha,\beta \ssmin \tilde{\Delta}}
              \Phi(\alpha(q),u) \, \Phi(\beta(q),v) \,
              \Bigl( i \, \mbox{\sl g}_\alpha \, \alpha(F_\beta) \, - \,
                     i \, \mbox{\sl g}_\beta \, \beta(F_\alpha) \Bigr) \;
              E_\alpha \otimes E_\beta                                       \\
 & &\!\! - \; {\textstyle {1 \over 2}} \,
              \sum_{\alpha,\beta \ssmin \tilde{\Delta}}
              i \, \mbox{\sl g}_\gamma \,
              \Bigl( \Phi(\alpha(q),u) \, \Phi(\beta(q),v) \,
                     N_{\alpha,\beta} \; E_\alpha \otimes E_\beta      \\[-6mm]
 & &\!\! \phantom{- \; {\textstyle {1 \over 2}} \,
                       \sum_{\alpha,\beta \ssmin \tilde{\Delta}}
                       i \, \mbox{\sl g}_\gamma \, \Bigl(}
                     - \, \Phi(\alpha(q),-u) \, \Phi(\beta(q),v) \,
                          N_{\alpha,\beta} \;
                          \theta E_\alpha \otimes E_\beta \Bigr)       \\[-3mm]
 &=&\!\! \sum_{\alpha \ssmin \tilde{\Delta}} i \, \mbox{\sl g}_\alpha \,
         \Bigl( \Phi^\prime(\alpha(q),v) \,
                (H_\alpha)_{\mathfrak{a}} \otimes E_\alpha \, - \,
                \Phi^\prime(\alpha(q),u) \,
                E_\alpha \otimes (H_\alpha)_{\mathfrak{a}} \,\Bigr)          \\
 & &\!\! + \; \sum_{\alpha,\beta \ssmin \tilde{\Delta}}
              \Phi(\alpha(q),u) \, \Phi(\beta(q),v) \,
              \Bigl( i \, \mbox{\sl g}_\alpha \, \alpha(F_\beta) \, - \,
                     i \, \mbox{\sl g}_\beta \, \beta(F_\alpha) \, - \,
                     i \, {\mit{\Gamma}}_{\alpha,\beta}^\theta \Bigr) \;
              E_\alpha \otimes E_\beta~.
\end{eqnarray*}
The first term gives precisely the {\em rhs} of eqn (\ref{eq:SERMAT3}),
so the last double sum must vanish. Thus we are led to conclude that, for
the elliptic Calogero model, the Poisson bracket relation (\ref{eq:ERMAT1})
will hold if and only if the following combinatoric identity is valid:
\begin{equation} \label{eq:SKOMBI2}
 \mbox{\sl g}_\alpha \, \alpha(F_\beta) \, - \,
 \mbox{\sl g}_\beta  \, \beta(F_\alpha)~
 =~{\mit{\Gamma}}_{\alpha,\beta}^\theta
 \qquad \mbox{for $\, \alpha,\beta \smin \tilde{\Delta}$}~.
\end{equation}
The same calculation with $\Phi(s,u)$ and $\Phi(s,v)$ both replaced by $w(s)$,
$\Phi(s,u-v)$ and $\Phi(s,v-u)$ both replaced by $\, - w^\prime(s)/w(s) \,$
and $\zeta$ replaced by zero leads to the conclusion that, for the ordinary
(rational or trigonometric) Calogero model, the Poisson bracket relation
(\ref{eq:ORMAT1}) will hold if and only if the same combinatoric identity
(\ref{eq:SKOMBI2}) is valid. Moreover, this identity is easily shown to be
equivalent to the previously imposed identity (\ref{eq:SKOMBI1}).

Once again, the two basic results thus obtained are not entirely independent
because the Hamiltonian is a quadratic function of the $L$-matrix whereas the
$A$-matrix is essentially just the composition of the $R$-matrix with the
$L$-matrix. More precisely,
\begin{equation} \label{eq:SORELHL}
 H~=~{\textstyle {1 \over 2}} \, (L,L)
\end{equation}
and, due to eqn ({\ref{eq:OFUNEQ3}),
\begin{equation} \label{eq:SORELRA}
 A~=~R \cdot L
\end{equation}
for the ordinary (rational or trigonometric) Calogero model while
\begin{equation} \label{eq:SERELHL}
 H~=~{\rm Res} \vert_{u=0}^{}
     \left( {1 \over 2u} \, (L(u),L(u)) \right)
\end{equation}
and, due to eqns ({\ref{eq:EFUNEQ2}) and (\ref{eq:EFUNEQ6}),
$$
 R(u,v) \cdot L(v)~
 =~A(u) \, - \,
   {\textstyle {1 \over 2}} \left( \zeta(u-v) + \zeta(u+v) \right) L(u) \, + \,
   i \, \wp(v) \, \sum_{\alpha \ssmin \tilde{\Delta}}
   \mbox{\sl g}_\alpha F_\alpha
$$
and hence
\begin{equation} \label{eq:SERELRA}
 A(u)~=~{\rm Res} \vert_{v=u}^{}
        \left( {1 \over v-u} \, R(u,v) \cdot L(v) \right) + \,
        {\textstyle {1 \over 2}} \, \zeta(2u) \, L(u)
\end{equation}
for the elliptic Calogero model, provided that the function $F$ mentioned
above satisfies the simple constraint equation
\begin{equation} \label{eq:SFUNF3}
 \sum_{\alpha \ssmin \tilde{\Delta}} \mbox{\sl g}_\alpha F_\alpha~=~0~.
\end{equation}
Thus once again, we may conclude that the Lax equations follow from the
Poisson bracket relations.

Concluding, we mention the fact that in view of the invariance of the
structure constants $N_{\alpha,\beta}$ under the action of the entire
Weyl group $W(\mathfrak{g})$ of $\mathfrak{g}$,
\begin{equation}
 N_{w\alpha,w\beta}~=~N_{\alpha,\beta} \qquad
 \mbox{for all $\, w \smin W(\mathfrak{g})$}~,
\end{equation}
together with the invariance of the coupling constants $\mbox{\sl g}_\alpha$
under the action of the subgroup $W_\theta(\mathfrak{g})$ of $W(\mathfrak{g})$
associated with the symmetric pair $(\mathfrak{g},\theta)$, as required in eqn
(\ref{eq:COUPC5}), the function $F$ must be covariant (equivariant) under the
action of this subgroup:
\begin{equation} \label{eq:SFUNF4}
 F_{w\alpha}~=~w(F_\alpha) \qquad
 \mbox{for all $\, w \smin W_\theta(\mathfrak{g})$}~.
\end{equation}

\subsection{Solution for the AIII-Series}

As an example that provides non-trivial solutions to the combinatoric identity
(\ref{eq:SKOMBI1}) derived above, we consider the symmetric pairs corresponding
to the complex Grass\-mannians, the symmetric spaces of the $AIII$-series,
which are given by the choice
\begin{equation} \label{eq:LIEALA3}
 \mathfrak{g}~=~\mathfrak{sl}(n,\mathbb{C})~~,~~
 \mathfrak{g}_0~=~\mathfrak{su}(p,q)~~,~~
 \mathfrak{g}_k~=~\mathfrak{su}(n)~,
\end{equation}
where $\, n = p+q \,$ and $\, p \leq q \,$. In order to take advantage of the
calculations performed in Sect.\ 3.2 for the $\mathfrak{sl}(n,\mathbb{C})$
case, we choose a representation in which the Cartan subalgebra $\mathfrak{h}$
of $\mathfrak{g}$ is the standard one, consisting of purely diagonal matrices;
this forces us to deviate from another standard convention, adopted, e.g.,
in ref.\ \cite{He}, according to which the involution $\theta$ is given by
conjugation with a matrix having $p$ entries equal to $+1$ and $q$ entries
equal to $-1$ on the diagonal. Instead, we shall represent all matrices in
$\mathfrak{sl}(n,\mathbb{C})$ in the $(3 \times 3)$ block form
\begin{equation} \label{eq:BLOCKA30}
 \begin{array}{rc}
  n~\raisebox{-0.3ex}{\Large $\updownarrow$} \!\!\!
                     & \Bigl( \ldots \Bigr) \\
                     & \longleftrightarrow  \\[-1ex]
                     & n
 \end{array}~\raisebox{2ex}{=}~
 \begin{array}{l}
  \left( \begin{array}{ccc}
          (\ldots) & \,(\ldots)\, & (\ldots) \\
          (\ldots) & \,(\ldots)\, & (\ldots) \\
          (\ldots) & \,(\ldots)\, & (\ldots)
         \end{array} \right) \!\!\!
         \begin{array}{cc}
          \updownarrow &  p  \\
          \updownarrow & q-p \\
          \updownarrow &  p
         \end{array}~, \\
  \hspace*{0.95em}
  \begin{array}{ccc}
   \phantom{(\ldots)}  & \phantom{(\ldots)}  & \phantom{(\ldots)}  \\[-3ex]
   \longleftrightarrow & \longleftrightarrow & \longleftrightarrow \\[-1ex]
            p          &         q-p         &          p          \\
  \end{array}
 \end{array}
\end{equation}
and the automorphism $\theta$ is given by conjugation
\begin{equation} \label{eq:INVOLA31}
 \theta X~=~J_{p,q} X J_{p,q} \qquad
 \mbox{for $\, X \smin \mathfrak{sl}(n,\mathbb{C})$}
\end{equation}
with the matrix
\begin{equation} \label{eq:INVOLA32}
 J_{p,q}~=~\left( \begin{array}{ccc}
                   0   &    0    & 1_p \\
                   0   & 1_{q-p} &  0  \\
                   1_p &    0    &  0
                \end{array} \right)~.
\end{equation}
Taking into account that the conjugation $\tau$ with respect to the compact
real form $\, \mathfrak{g}_k = \mathfrak{su}(n) \,$ should be given by
\begin{equation} \label{eq:CONJA31}
 \tau X  ~=~- \, X^\dagger \qquad
 \mbox{for $\, X \smin \mathfrak{sl}(n,\mathbb{C})$}~,
\end{equation}
where $.^\dagger$ denotes hermitean adjoint, in order to guarantee that the
matrices constituting $\, \mathfrak{g}_k = \mathfrak{su}(n) \,$ continue to be
the traceless antihermitean matrices, we conclude that the conjugation $\sigma$
with respect to the real form $\, \mathfrak{g}_0 = \mathfrak{su}(p,q) \,$
must in this representation be given by
\begin{equation} \label{eq:CONJA32}
 \sigma X~=~- \, J_{p,q} X^\dagger J_{p,q} \qquad
 \mbox{for $\, X \smin \mathfrak{sl}(n,\mathbb{C})$}~,
\end{equation}
so the matrices constituing $\, \mathfrak{g}_0 = \mathfrak{su}(p,q) \,$ are
the traceless matrices of the form
\begin{equation} \label{eq:BLOCKA31}
 \left( \begin{array}{ccc}
         A &      C      &   E         \\
         D &      B      & - C^\dagger \\
         F & - D^\dagger & - A^\dagger
        \end{array} \right)~,
\end{equation}
where $A$, $C$ and $D$ are arbitrary complex matrices whereas $B$, $E$ and
$F$ are antihermitean matrices. Similarly, the matrices constituing the
intersection $\mathfrak{k}_0$ of the two real forms are the traceless
matrices of the form
\begin{equation} \label{eq:BLOCKA32}
 \left( \begin{array}{ccc}
              A_1      & C_1 &      E_1      \\
         - C_1^\dagger & B_1 & - C_1^\dagger \\
              E_1      & C_1 &      A_1
        \end{array} \right)~,
\end{equation}
where $C_1$ is an arbitrary complex matrix whereas $A_1$, $B_1$ and $E_1$ are
antihermitean matrices. Finally, the matrices belonging to the complementary
space $\mathfrak{m}_0$ appearing in the decompositions (\ref{eq:CARDEC2}) and
(\ref{eq:CARDEC3}) are the traceless matrices of the form
\begin{equation} \label{eq:BLOCKA33}
 \left( \begin{array}{ccc}
            A_2      &  C_2  &      E_2 \\
         C_2^\dagger &   0   & - C_2^\dagger \\
           - E_2     & - C_2 &     - A_2
        \end{array} \right)~,
\end{equation}
where $C_2$ is an arbitrary complex matrix whereas $A_2$ is a hermitean matrix
and $E_2$ is an antihermitean matrix. In particular, the $p$-dimensional sub%
space $\mathfrak{a}_0$ that appears as the ambient space for the configuration
space of the Calogero models consists of the real diagonal matrices of the form
\begin{equation} \label{eq:BLOCKA34}
 \left( \begin{array}{ccc}
         P & 0 &  0 \\
         0 & 0 &  0 \\
         0 & 0 & - P
        \end{array} \right)~,
\end{equation}
whereas its $(q-1)$-dimensional orthogonal complement $i\mathfrak{b}_0$ in
$\mathfrak{h}_{\mathbb{R}}$ consists of the real traceless diagonal matrices
of the form
\begin{equation} \label{eq:BLOCKA35}
 \left( \begin{array}{ccc}
         P & 0 & 0 \\
         0 & Q & 0 \\
         0 & 0 & P
        \end{array} \right)~.
\end{equation}
For later use, we also introduce the diagonal $(n \times n)$-matrices
\begin{equation} \label{eq:BLOCKA36}
 I_{2p}~=~\left( \begin{array}{ccc}
                  1_p & 0 &  0 \\
                   0  & 0 &  0 \\
                   0  & 0 & 1_p
                 \end{array} \right)\!~~,~~
 I_q~=~\left( \begin{array}{ccc}
               0 &  0  & 0 \\
               0 & 1_q & 0 \\
               0 &  0  & 0
              \end{array} \right)~.
\vspace{2mm}
\end{equation}
Passing to the explicit index calculations, we continue to use the notation
and the conventions introduced in Sect.\ 3.2 for handling the $\mathfrak{sl}%
(n,\mathbb{C})$ case. In particular, we continue to let indices $\, a,b,
\ldots \,$ run from $1$ to $n$, but we shall use the following terminology
to characterize the subdivision of this range indicated by the $(3 \times 3)$
block form introduced above: an index $a$ will be said to belong to the first
block if $\, 1 \leq a \leq p$, to the second block if $\, p+1 \leq a \leq q \,$
and to the third block if $\, q+1 \leq a \leq n$; moreover, we introduce the
following abbreviation to characterize the action of the involution $\theta$
on these indices:
\vspace{5mm}
\begin{equation} \label{eq:INVOLA33}
 \begin{array}{ccc}
  \theta(a)~=~a+q & \quad & \mbox{for $\, 1 \leq a \leq p$} \\[1mm]
  \theta(a)~=~a   & \quad & \mbox{for $\, p+1 \leq a \leq q$} \\[1mm]
  \theta(a)~=~a-q & \quad & \mbox{for $\, q+1 \leq a \leq n$}
 \end{array}~.
\vspace{2mm}
\end{equation}
This allows for a considerable simplification of the notation. For example,
we have
\begin{equation} \label{eq:INVOLA34}
 \theta \alpha_{ab}~=~\alpha_{\theta(a)\theta(b)}~,
\vspace{-1mm}
\end{equation}
with
\begin{equation} \label{eq:INVOLA35}
 \theta E_{ab}~=~E_{\theta(a)\theta(b)}~.
\vspace{2mm}
\end{equation}
In particular, the two parts of the root system $\Delta$ of $\mathfrak{sl}%
(n,\mathbb{C})$ appearing in the decomposition (\ref{eq:SDECRS}) are given by
\begin{equation}
 \tilde{\Delta}~
 =~\{ \, \alpha_{ab} \smin \Delta \, / \,
         \mbox{$a$ or $b$ belongs to the first or third block} \, \}~,
\end{equation}
and
\begin{equation}
 \Delta_0~
 =~\{ \, \alpha_{ab} \smin \Delta \, / \,
         \mbox{$a$ and $b$ belong to the second block} \, \}~.
\vspace{2mm}
\end{equation}
For the sake of completeness and for later use, we specify explicitly
the elements of $\tilde{\Delta}$ and the resulting restricted root system
$\bar{\Delta}$, with the corresponding multiplicities. Introducing an
orthonormal basis $\, \{ e_1 , \ldots , e_p \} \,$ of $\mathfrak{a}_0^\ast$
by
\begin{equation} \label{eq:RSBAS3}
 e_k \left( \begin{array}{ccc}
             P & 0 &  0 \\
             0 & 0 &  0 \\
             0 & 0 & - P
            \end{array} \right)\!~=~P_{kk} \qquad
 \mbox{for} \, \left( \begin{array}{ccc}
                       P & 0 &  0 \\
                       0 & 0 &  0 \\
                       0 & 0 & - P
                      \end{array} \right) \smin \, \mathfrak{a}_0~,
\end{equation}
we have
\begin{equation} \label{eq:RRSA3}
 \begin{array}{ccc}
  \bar{\alpha}_{kl}~=~\bar{\alpha}_{\theta(l)\theta(k)}~=~e_k - e_l &
  (1 \leq k < l \leq p) & \mbox{multiplicity $2$}                       \\[1mm]
  \bar{\alpha}_{lk}~=~\bar{\alpha}_{\theta(k)\theta(l)}~=~e_l - e_k &
  (1 \leq k < l \leq p) & \mbox{multiplicity $2$}                       \\[2mm]
  \bar{\alpha}_{k\,\theta(l)}~=~\bar{\alpha}_{l\,\theta(k)}~=~+ \, e_k + e_l &
  (1 \leq k < l \leq p) & \mbox{multiplicity $2$}                       \\[1mm]
  \bar{\alpha}_{\theta(k)\,l}~=~\bar{\alpha}_{\theta(l)\,k}~=~- \, e_k - e_l &
  (1 \leq k < l \leq p) & \mbox{multiplicity $2$}                       \\[2mm]
  \bar{\alpha}_{k\,\theta(k)}~=~+ \, 2 \>\! e_k &
  (1 \leq k \leq p) & \mbox{multiplicity $1$}                           \\[1mm]
  \bar{\alpha}_{\theta(k)\,k}~=~- \, 2 \>\! e_k &
  (1 \leq k \leq p) & \mbox{multiplicity $1$}                           \\[2mm]
  \bar{\alpha}_{km}~=~\bar{\alpha}_{m\,\theta(k)}~=~+ \, e_k & \left\{
  \begin{array}{c}
   (1 \leq k \leq p) \\
   (p+1 \leq m \leq q)
  \end{array} \right\} & \mbox{multiplicity $2(q-p)$}                   \\[4mm]
  \bar{\alpha}_{mk}~=~\bar{\alpha}_{\theta(k)\,m}~=~- \, e_k & \left\{
  \begin{array}{c}
   (1 \leq k \leq p) \\
   (p+1 \leq m \leq q)
  \end{array} \right\} & \mbox{multiplicity $2(q-p)$}
 \end{array}
\end{equation}
The corresponding root generators continue to be given by $\, E_{\alpha_{ab}}
\equiv E_{ab} \,$ ($1 \leq a \neq b \leq n$, with $a$ or $b$ belonging to the
first or third block), since these also satisfy the constraints listed in eqn
(\ref{eq:NORM3}). However, in contrast to the $\mathfrak{sl}(n,\mathbb{C})$
case studied in Sect.\ 3.2, we must now deal with the possibility of
encountering nontrivial coupling constants $\, \mbox{\sl g}_{\alpha_{ab}}
\equiv \mbox{\sl g}_{ab}$ ($1 \leq a,b \leq n$). A priori, these are only
defined when $a$ or $b$ belong to the first or third block, but their
definition will be extended to all values of $a$ and $b$ by setting
\begin{equation} \label{eq:COUPC6}
 \mbox{\sl g}_{ab}~=~0 \qquad
 \mbox{if $a$ and $b$ belong to the second block}~,
\end{equation}
as postulated in to eqn (\ref{eq:COUPC3}), and by adding the convention
\begin{equation} \label{eq:COUPC7}
 \mbox{\sl g}_{aa}~=~0~.
\end{equation}
Using eqn (\ref{eq:STRCNA}), we can then derive the following simple
expressions for the combinations \linebreak $\, {\mit{\Gamma}}_%
{\alpha_{ab},\alpha_{cd}}^{} \equiv {\mit{\Gamma}}_{ab,cd}^{} \,$
and $\, {\mit{\Gamma}}_{\alpha_{ab},\alpha_{cd}}^\theta \equiv
{\mit{\Gamma}}_{ab,cd}^\theta \,$ introduced in eqns (\ref{eq:GAMMA})
and (\ref{eq:SGAMMA}), respectively:
\begin{equation} \label{eq:GAMMAA3}
 {\mit{\Gamma}}_{ab,cd}^{}~
 =~\mbox{\sl g}_{ad} \, \delta_{bc}^{} \, - \,
   \mbox{\sl g}_{bc} \, \delta_{ad}^{}~,
\vspace{-5mm}
\end{equation}
\begin{eqnarray} \label{eq:SGAMMAA3}
 \begin{array}{rcl}
  {\mit{\Gamma}}_{ab,cd}^\theta \!\!
  &=&\!\! {\textstyle {1 \over 4}}
          \Bigl( \mbox{\sl g}_{ad} \, \delta_{bc}^{} \,
            - \, \mbox{\sl g}_{bc} \, \delta_{ad}^{} \,
            + \, \mbox{\sl g}_{\theta(a)\,d} \, \delta_{\theta(b)\,c}^{} \,
            - \, \mbox{\sl g}_{\theta(b)\,c} \, \delta_{\theta(a)\,d}^{}
  \\[-1mm]
  & &\!\! \phantom{{\textstyle {1 \over 4}} \Bigl(}
            + \, \mbox{\sl g}_{a\,\theta(d)} \, \delta_{b\,\theta(c)}^{} \,
            - \, \mbox{\sl g}_{b\,\theta(c)} \, \delta_{a\,\theta(d)}^{} \,
            + \, \mbox{\sl g}_{\theta(a)\theta(d)} \,
                 \delta_{\theta(b)\theta(c)}^{} \,
            - \, \mbox{\sl g}_{\theta(b)\theta(c)} \,
                 \delta_{\theta(a)\theta(d)}^{} \Bigr)~,
 \end{array}&&
\end{eqnarray}
valid for $\, 1 \leq a,b,c,d \leq n \,$ with $\, a \neq b \,$ and
$\, c \neq d$. Next, we have for the Weyl reflections
\begin{equation} \label{eq:INVOLA36}
 \theta s_{ab} \, \theta~=~s_{\theta(a)\theta(b)}~.
\end{equation}
This relation implies that the Weyl group $W(\mathfrak{sl}(n,\mathbb{C}),
\theta)$ associated with this symmetric pair can be realized not only as
the quotient of a subgroup but actually as a subgroup of the Weyl group
$W(\mathfrak{sl}(n,\mathbb{C}))$ of $\mathfrak{sl}(n,\mathbb{C})$,
namely the subgroup generated by the reflections $\, s_{a\,\theta(a)}
= s_{\theta(a)\,a} \,$ with index $a$ in the first block (or equivalently
in the third block) and by the products of reflections $\, s_{ab} \,
s_{\theta(a)\theta(b)} = s_{\theta(a)\theta(b)} \, s_{ab} \,$ with
both indices $a$ and $b$ in the first block (or equivalently in the
third block). Under the action of this group, $\tilde{\Delta}$
decomposes into three distinct orbits, namely
\begin{eqnarray}
 &\{ \, \alpha_{ab} \, / \,
        \mbox{$a$ and $b$ belong to the first or third block,
        $b \neq a$, $b \neq \theta(a)$} \, \}~, \quad &                 \\[3mm]
 &\{ \, \alpha_{a\,\theta(a)} \, / \,
        \mbox{$a$ belongs to the first or third block} \, \}~, \quad &  \\[5mm]
 &\Big\{ \, \alpha_{ab}~\Big/~
            \begin{minipage}{11cm}
             \begin{center}
              $a$ belongs to the first or third block
              and $b$ to the second block \\[-0.5ex]
              or \\[-0.5ex]
              $a$ belongs to the second block
              and $b$ to the first or third block
             \end{center}
            \end{minipage}
            \, \Big\}~, \quad &
\end{eqnarray}
and so there will be three independent coupling constants which, following
refs \cite{OP1,OP2,Pe}, we shall denote by $\mbox{\sl g}$, $\mbox{\sl g}_1$
and $\mbox{\sl g}_2$. Explicitly,

\pagebreak

\begin{itemize}
 \item $\mbox{\sl g}_{ab} = \mbox{\sl g} \,$ if $a$ and $b$ belong to the
       first or third block, with $\, b \neq a \,$ and $\, b \neq \theta(a)$,
 \item $\mbox{\sl g}_{a\,\theta(a)} = \mbox{\sl g}_{\theta(a)\,a}
       = \mbox{\sl g}_2 \,$ if $a$ belongs to the first or third block,
 \item $\mbox{\sl g}_{ab} = \mbox{\sl g}_1 \,$ if $a$ belongs to the first
       or third block and $b$ to the second block or if $a$ belongs to the
       second block and $b$ to the first or third block.
\end{itemize}
Note also that at least the coupling constant $\mbox{\sl g}$ should be non-%
zero, since otherwise, the Hamiltonian of the corresponding Calogero model
(as given in eqns (\ref{eq:SOHAMF1}), (\ref{eq:SOHAMF2}) or (\ref{eq:SEHAMF1}),
(\ref{eq:SEHAMF2})) would decouple, that is, would decompose into the sum of
$p$ copies of the same Hamiltonian for a system with only one degree of
freedom, and such a system is trivially completely integrable.

With these preliminaries out of the way, we proceed to search for a collection
of matrices $\, F_{\alpha_{ab}} \equiv F_{ab} \,$ ($1 \leq a \neq b \leq n$,
with $a$ or $b$ belonging to the first or third block) that belong to
$i\mathfrak{b}_0$, with $\, F_{ba} = F_{ab} = F_{\theta(a)\theta(b)} \,$ as
required in eqn (\ref{eq:SFUNF2}), satisfying the property (\ref{eq:SFUNF4})
of covariance under the Weyl group and the combinatoric identity
\begin{eqnarray} \label{eq:SKOMBI3}
\lefteqn{2 \, \mbox{\sl g}_{cd}
         \left( (F_{ab})_{cc} - (F_{ab})_{dd} \right)\!~
         =~2 \, \mbox{\sl g}_{cd} \, \alpha_{cd}(F_{ab})~
         =~{\mit{\Gamma}}_{cd,ab}^\theta + {\mit{\Gamma}}_{cd,ba}^\theta}
                                                      \nonumber \hspace{3mm} \\
 &=&\!\! {\textstyle {1 \over 4}} \,
         \Bigl( \mbox{\sl g}_{cb} \, \delta_{da}^{} \, - \,
                \mbox{\sl g}_{da} \, \delta_{cb}^{} \, + \,
                \mbox{\sl g}_{ca} \, \delta_{db}^{} \, - \,
                \mbox{\sl g}_{db} \, \delta_{ca}^{}                    \\[-3mm]
 & &\!\! \phantom{{\textstyle {1 \over 4}} \, \Bigl(}
                + \, \mbox{\sl g}_{\theta(c)\,b} \, \delta_{\theta(d)\,a}^{} \,
                - \, \mbox{\sl g}_{\theta(d)\,a} \, \delta_{\theta(c)\,b}^{} \,
                + \, \mbox{\sl g}_{\theta(c)\,a} \, \delta_{\theta(d)\,b}^{} \,
                - \, \mbox{\sl g}_{\theta(d)\,b} \, \delta_{\theta(c)\,a}^{}
                                                             \nonumber \\[-3mm]
 & &\!\! \phantom{{\textstyle {1 \over 4}} \, \Bigl(}
                + \, \mbox{\sl g}_{c\,\theta(b)} \, \delta_{d\,\theta(a)}^{} \,
                - \, \mbox{\sl g}_{d\,\theta(a)} \, \delta_{c\,\theta(b)}^{} \,
                + \, \mbox{\sl g}_{c\,\theta(a)} \, \delta_{d\,\theta(b)}^{} \,
                - \, \mbox{\sl g}_{d\,\theta(b)} \, \delta_{c\,\theta(a)}^{}
                                                             \nonumber \\[-3mm]
 & &\!\! \phantom{{\textstyle {1 \over 4}} \, \Bigl(}
                + \, \mbox{\sl g}_{\theta(c)\theta(b)} \,
                     \delta_{\theta(d)\theta(a)}^{} \,
                - \, \mbox{\sl g}_{\theta(d)\theta(a)} \,
                     \delta_{\theta(c)\theta(b)}^{} \,
                + \, \mbox{\sl g}_{\theta(c)\theta(a)} \,
                     \delta_{\theta(d)\theta(b)}^{} \,
                - \, \mbox{\sl g}_{\theta(d)\theta(b)} \,
                     \delta_{\theta(c)\theta(a)}^{} \Bigr)~, \quad \nonumber
\end{eqnarray}
valid for $\, 1 \leq a,b,c,d \leq n \,$ with $\, a \neq b \,$ and
$\, c \neq d$, corresponding to eqn (\ref{eq:SKOMBI1}). Note that
Note also that
the condition $\, F_{ab} \smin i\mathfrak{b}_0 \,$ means that
$F_{ab}$ must be a real traceless diagonal matrix whose entries
satisfy
\begin{equation} \label{eq:SFUNF5}
 (F_{ab})_{cc}~=~(F_{ab})_{\theta(c)\theta(c)}~,
\end{equation}
whereas the invariance under the reflections $\, s_{a\,\theta(a)}
= s_{\theta(a)\,a} \,$ with index $a$ in the first block (or
equivalently in the third block), which belong to the Weyl group
$W(\mathfrak{sl}(n,\mathbb{C}),\theta)$ and act trivially on the
space $i\mathfrak{b}_0$, means that
\begin{equation} \label{eq:SFUNF6}
 F_{ab}~=~F_{\theta(a)\,b}~=~F_{a\,\theta(b)}~=~F_{\theta(a)\theta(b)}~.
\end{equation}
Therefore, it suffices to determine $F_{ab}$ in the following three cases:
when $a$ and $b$ both belong to the first block, with $\, a \neq b$, when
$a$ belongs to the first block while $b$ belongs to the third block, with
$\, b = \theta(a)$, and finally when $a$ belongs to the first block while
$b$ belongs to the second block. Moreover, the system of equations
(\ref{eq:SKOMBI3}) is obviously invariant under the substitution
$\, c \rightarrow \theta(c)$, $d \rightarrow \theta(d) \,$ and under
the exchange of $c$ and $d$, as well as under that of $a$ and $b$.
Now obviously, the {\em rhs} of eqn (\ref{eq:SKOMBI3}) vanishes if the
sets $\{c,d\}$ and $\{a,b,\theta(a),\theta(b)\}$ are disjoint, so eqn
(\ref{eq:SKOMBI3}) will in this case be satisfied, independently of the
choice of the coupling constants, if we assume all entries of the matrix
$F_{ab}$ except $(F_{ab})_{aa}$, $(F_{ab})_{bb}$, $(F_{ab})_{\theta(a)%
\theta(a)}$ and $(F_{ab})_{\theta(b)\theta(b)}$ to be equal among themselves.
Similarly, the case when the set $\{c,d\}$ is contained in the set $\{a,b,%
\theta(a),\theta(b)\}$ will provide relations between the remaining entries
of the matrix $F_{ab}$. \linebreak To find these, we observe that, due to the invariance
mentioned above, it is in this case sufficient to evaluate eqn (\ref{eq:%
SKOMBI3}) with $\, c = a \,$ and $\, d = b$, $d = \theta(b)$, $d = \theta(a)$.
Now when $a$ and $b$ both belong to the first block, with $\, a \neq b$, the
first two cases ($d = b \,$ and $\, d = \theta(b)$) lead to the relation
$$
 2 \, \mbox{\sl g} \left( (F_{ab})_{aa} - (F_{ab})_{bb} \right)\!~=~0~,
$$
while the third ($d = \theta(a)$) gives a trivial identity. Similarly, when
$a$ belongs to the first block while $b$ belongs to the third block, with
$\, b = \theta(a)$, the first and last case coincide ($d = b = \theta(a)$)
and give a trivial identity, while the second ($d = \theta(b)$) is excluded.
Thus taking into account the fact that $\, \mbox{\sl g} \neq 0$, we can cover
both situations in one stroke by writing
$$
 F_{ab}~=~{1 \over 4} \, \lambda_{ab}
          \left( E_{aa} + E_{\theta(a)\theta(a)} +
                 E_{bb} + E_{\theta(b)\theta(b)} \right) + \,
          {1 \over n} \, \tau_{ab} \, 1
$$
with real coefficients $\lambda_{ab}$ and $\tau_{ab}$ to be determined.
Finally, when $a$ belongs to the first block while $b$ belongs to the
second block, the first two cases coincide ($d = b = \theta(b)$) and
lead to the relation
$$
 2 \, \mbox{\sl g}_1 \left( (F_{ab})_{aa} - (F_{ab})_{bb} \right)\!~
 =~{\textstyle {1 \over 2}} \, \mbox{\sl g}_2
$$
while the third ($d = \theta(a)$) gives a trivial identity, suggesting
that we should write
$$
 F_{ab}~=~{1 \over 2} \, \lambda_{ab}
          \left( E_{aa} + E_{\theta(a)\theta(a)} \right) + \,
          \mu_{ab} \, E_{bb} \, + \, {1 \over n} \, \tau_{ab} \, 1
$$
with real coefficients $\lambda_{ab}$, $\mu_{ab}$ and $\tau_{ab}$ to be
determined, subject to the constraint
$$
 2 \, \mbox{\sl g}_1 \left( \lambda_{ab} - 2 \>\! \mu_{ab} \right)\!~
 =~\mbox{\sl g}_2~.
$$
It remains to check eqn (\ref{eq:SKOMBI3}) for the cases where the intersection
of the sets $\{c,d\}$ and $\{a,b,\theta(a),\theta(b)\}$ contains precisely one
element. Once again, due to the invariance mentioned above, it is in this case
sufficient to evaluate eqn (\ref{eq:SKOMBI3}) with $\, c = a \,$ and with
$\, c = b$, supposing that $\, d \nsmin \{a,b,\theta(a),\theta(b)\}$.
When $a$ and $b$ both belong to the first block and also when $a$ belongs to
the first block while $b$ belongs to the third block, with $\, b = \theta(a)$,
this equation is solved identically, independently of the choice of the
coupling constants, by putting $\, \lambda_{ab} = -1$. On the other hand,
when $a$ belongs to the first block while $b$ belongs to the second block,
we obtain non-trivial relations between the coupling constants, which are the
following. If $\, c = a \neq b$, choosing $\, d \nsmin \{a,b,\theta(a)\} \,$
to belong to the first or third block gives
$$
 2 \, \mbox{\sl g} \, \lambda_{ab}~=~- \, \mbox{\sl g}_1~,
$$
while choosing $\, d \nsmin \{a,b,\theta(a)\} \,$ to belong to the second block
(which is only possible when $\, q \geq p+2$) gives
$$
 \mbox{\sl g}_1 \, \lambda_{ab}~=~0~.
$$
Similarly, if $\, c = b \neq a$, choosing $\, d \nsmin \{a,b,\theta(a)\} \,$
to belong to the first or third block gives
$$
 2 \, \mbox{\sl g}_1 \, \mu_{ab}~=~- \, \mbox{\sl g}~,
$$
while choosing $\, d \nsmin \{a,b,\theta(a)\} \,$ to belong to the second block
(which is only possible when $\, q \geq p+2$) gives
$$
 \mbox{\sl g}_1~=~0~.
$$
Finally, fixing the coefficients $\tau_{ab}$ by requiring that $F_{ab}$ should
be traceless, we arrive at the following
\begin{theorem}
 For the symmetric pairs associated with the complex Grassmannians, which
 are the symmetric spaces of the $AIII$-series corresponding to the choice
 (\ref{eq:LIEALA3}) with $\, n = p+q \,$ and $\, p \leq q \,$, a non-trivial
 solution to the combinatoric identity (\ref{eq:SKOMBI1}) exists if and only
 if either $\, q = p \,$ or $\, q = p+1 \,$ and is given as follows:
 \begin{itemize}
  \item $q = p \,$: \\[1mm]
        The function $F$ is given by
        \begin{equation} \label{eq:SFUNF7}
         F_{ab}~=~- \, {1 \over 4}
                       \left( E_{aa} + E_{\theta(a)\theta(a)} +
                              E_{bb} + E_{\theta(b)\theta(b)} \right)
                  + \, {1 \over n} \, 1
        \end{equation}
  \item $q = p+1 \,$: \\[1mm]
        The function $F$ is given by
        \begin{equation} \label{eq:SFUNF8}
         F_{ab}~=~- \, {1 \over 4}
                       \left( E_{aa} + E_{\theta(a)\theta(a)} +
                              E_{bb} + E_{\theta(b)\theta(b)} \right)
                  + \, {1 \over n} \, 1
        \end{equation}
        when $a$ and $b$ both belong to the first or third block and by
        \begin{equation} \label{eq:SFUNF9}
         F_{ab}~=~- \, {\mbox{\sl g}_1 \over 4 \>\! \mbox{\sl g}}
                       \left( E_{aa} + E_{\theta(a)\theta(a)} \right)
                  - \, {\mbox{\sl g} \over 2 \>\! \mbox{\sl g}_1} \, E_{bb} \,
                  + \, {1 \over n}
                       \left( {\mbox{\sl g}_1 \over 2 \>\! \mbox{\sl g}} \, + \,
                              {\mbox{\sl g} \over 2 \>\! \mbox{\sl g}_1} \right)
                       1
        \end{equation}
        when $a$ belongs to the first or third block while $b$ belongs to the
        second block, provided that the coupling constants $\mbox{\sl g}$,
        $\mbox{\sl g}_1$ and $\mbox{\sl g}_2$ satisfy the relations
        \begin{equation} \label{eq:COUPC8}
         \mbox{\sl g}~\neq~0~~,~~\mbox{\sl g}_1~\neq~0~,
        \end{equation}
        together with
        \begin{equation} \label{eq:COUPC9}
         \mbox{\sl g}_1^2 \, - \, 2 \>\! \mbox{\sl g}_{}^2 \, + \,
         \mbox{\sl g} \>\! \mbox{\sl g}_2~=~0~.
        \end{equation}
 \end{itemize}
 This solution also satisfies the constraint equation (\ref{eq:SFUNF3}).
\end{theorem}
The only statement that has not yet been proved is the last one, referring
to the identity (\ref{eq:SFUNF3}). This is easily done by computing separately
the sum over the different Weyl group orbits: Using the matrices introduced in
eqn (\ref{eq:BLOCKA36}), we see that the first gives $4$ times a contribution
of the form
$$
 \mbox{\sl g} \sum_{a,b=1 \atop a \neq b}^p F_{ab}~
 =~\mbox{\sl g} \left( - \, {p-1 \over 2} \, I_{2p} \, + \,
                       {p \>\! (p-1) \over n} \, 1 \right)\!~,
$$
while the second gives $2$ times a contribution of the form
$$
 \mbox{\sl g}_2 \sum_{a=1}^p F_{a\,\theta(a)}~
 =~\mbox{\sl g}_2 \left( - \, {1 \over 2} \, I_{2p} \, + \,
                       {p \over n} \, 1 \right)\!~,
$$
and finally the third (which only exists if $\, q > p$) gives $4$ times a
contribution of the form
$$
 \mbox{\sl g}_1 \sum_{a=1}^p \sum_{b=1}^{q-p} F_{ab}~
 =~\mbox{\sl g}_1 \left( - \, {\mbox{\sl g}_1 \over 4 \>\! \mbox{\sl g}} \,
                              (q-p) \, I_{2p} \,
                         - \, {\mbox{\sl g} \over 2 \>\! \mbox{\sl g}_1} \,
                              p \, I_q \,
                         + \, {1 \over n} \left(
                              {\mbox{\sl g}_1 \over 2 \>\! \mbox{\sl g}} \,
                         + \, {\mbox{\sl g} \over 2 \>\! \mbox{\sl g}_1}
                              \right) p \>\! (q-p) \, 1 \right)\!~.
$$
Summing up everything and using that $q-p$ is either $0$ or $1$ and that in
the second case, eqn (\ref{eq:COUPC9}) has to be taken into account, we get
zero, as desired.

To conclude this section, it seems worthwhile to briefly discuss a few
limiting cases, noting that the vanishing of certain coupling constants
allows to effectively reduce the restricted root system to a subsystem.
In fact, eqn (\ref{eq:RRSA3}) reveals that the system of restricted
roots associated with the complex Grassmannian $SU(n,n)/S(U(n) \times
U(n))$ and with the complex Grassmannian $SU(n,n+1)/S(U(n) \times
U(n+1))$ is of type $C_n$ and of type $BC_n$, respectively.%
\footnote{We now use $n$ instead of $p$, as before.} Inserting
into the Hamiltonian (\ref{eq:SOHAMF1}) or (\ref{eq:SEHAMF1})
and using the abbreviation (\ref{eq:POTF}) for the potential,
we obtain the following special cases:
\begin{itemize}
 \item $SU(n,n)/S(U(n) \times U(n))$, $\mbox{\sl g}_2 = 0 \,$:
       \begin{equation} \label{eq:HAMFD}
        H(q,p)~=~{\textstyle {1 \over 2}} \, \sum_{k=1}^n \, p_k^2 \, + \,
                 \mbox{\sl g}^2 \! \sum_{1 \leq k \neq l \leq n} \!
                 \left( V(q_k - q_l) + V(q_k + q_l) \right)~~~
       \end{equation}
       which is the Calogero Hamiltonian associated to the root system $D_n$.
 \item $SU(n,n)/S(U(n) \times U(n))$, $\mbox{\sl g}_2$ arbitrary:
       \begin{equation} \label{eq:HAMFC}
        H(q,p)~=~{\textstyle {1 \over 2}} \, \sum_{k=1}^n \, p_k^2 \, + \,
                 \mbox{\sl g}^2 \! \sum_{1 \leq k \neq l \leq n} \!
                 \left( V(q_k - q_l) + V(q_k + q_l) \right) \, + \,
                 \mbox{\sl g}_2^2 \, \sum_{k=1}^n \, V(2 \;\! q_k)
       \end{equation}
       which is the Calogero Hamiltonian associated to the root system $C_n$.
 \item $SU(n,n+1)/S(U(n) \times U(n+1))$, $\mbox{\sl g}_2 = 0 \,$:
       \begin{equation} \label{eq:HAMFB}
        H(q,p)~=~{\textstyle {1 \over 2}} \, \sum_{k=1}^n \, p_k^2 \, + \,
                 \mbox{\sl g}^2 \! \sum_{1 \leq k \neq l \leq n} \!
                 \left( V(q_k - q_l) + V(q_k + q_l) \right) \, + \,
                 2 \;\! \mbox{\sl g}_1^2 \, \sum_{k=1}^n \, V(q_k)
       \end{equation}
       which is the Calogero Hamiltonian associated to the root system
       $B_n$, subject to the constraint imposed by the quadratic equation
       (\ref{eq:COUPC9}), which in this case reduces to $\, \mbox{\sl g}_1^2
       = 2 \>\! \mbox{\sl g}^2$.
 \item $SU(n,n+1)/S(U(n) \times U(n+1))$, $\mbox{\sl g}_2$ arbitrary
       (the general case):
       \begin{equation} \label{eq:HAMFBC}
        \begin{array}{rcl}
         H(q,p) \!\!
         &=&\!\! {1 \over 2} \, {\displaystyle {\sum_{k=1}^n \, p_k^2 \, + \,
                 \mbox{\sl g}^2 \! \sum_{1 \leq k \neq l \leq n} \!
                 \left( V(q_k - q_l) + V(q_k + q_l) \right)}}           \\[5mm]
         & &\!\! \phantom{{1 \over 2} \,
                          {\displaystyle {\sum_{k=1}^n \, p_k^2 \,}}}
                 {\displaystyle {\mbox{}
                 + \, \mbox{\sl g}_2^2 \, \sum_{k=1}^n \, V(2 \>\! q_k) \,
                 + \, 2 \;\! \mbox{\sl g}_1^2 \, \sum_{k=1}^n \, V(q_k)}}
        \end{array} \quad
       \end{equation}
       which is the Calogero Hamiltonian associated to the non-reduced root
       system $BC_n$, subject to the constraint imposed by the quadratic
       equation (\ref{eq:COUPC9}), which in particular requires
       $\, \mbox{\sl g}_1 \neq 0$.
\end{itemize}
Thus all Calogero Hamiltonians associated with the classical root systems
are completely integrable, admitting a Lax pair formulation with a dynamical
$R$-matrix.

\end{document}